\documentclass{article}

\usepackage[utf8]{inputenc}
\usepackage{graphicx, amsmath}
\usepackage{mathtools}
\DeclarePairedDelimiter\bra{\langle}{\rvert}
\DeclarePairedDelimiter\ket{\lvert}{\rangle}
\DeclarePairedDelimiterX\braket[2]{\langle}{\rangle}{#1\,\delimsize\vert\,\mathopen{}#2}

\usepackage[left=2cm, right=2cm, top=2cm]{geometry}

\usepackage{graphicx}
\usepackage{tabularx}
\usepackage{tabu}
\usepackage{longtable}
\usepackage{hyperref}
\usepackage{siunitx}

\usepackage{authblk}
\usepackage{subcaption}
\usepackage[export]{adjustbox} 

\makeatletter
\renewcommand\@cite[2]{\textsuperscript{#1}}
\makeatother
\usepackage[normalem]{ulem}

\usepackage{chemformula}
\setchemformula{kroeger-vink=true}

\DeclareSIUnit \units {units}

\title{\textbf{Probing cellular activity via charge-sensitive quantum nanoprobes}} 

\author[1]{Uri Zvi}
\author[2]{Shivam Mundhra}
\author[1]{David Ovetsky}
\author[1]{Qing Chen}
\author[2]{Aidan R. Jones}
\author[2]{Stella Wang}
\author[3]{Maria Roman}
\author[4]{Michele Ferro} 
\author[5]{Kunle Odunsi} 
\author[4]{Marina C. Garassino} 
\author[6,7]{Michael E. Flatté}
\author[1]{Melody Swartz} 
\author[6]{Denis R. Candido~\textsuperscript{\textdagger}}
\author[1,8]{Aaron Esser-Kahn~\textsuperscript{\textdagger}}
\author[1,8,9]{Peter C. Maurer~\textsuperscript{\textdagger}}
\affil[1]{\small Pritzker School of Molecular Engineering, University of Chicago,
Chicago, IL 60637, USA.}
\affil[2]{\small Department of Physics, University of Chicago, Chicago, IL 60637, USA.}
\affil[3]{\small Department of Chemistry, University of Chicago, Chicago, IL 60637, USA.}
\affil[4]{\small Section of Hematology/Oncology, Department of Medicine, The University of Chicago, Chicago, Illinois, USA.}
\affil[5]{\small University of Chicago Medicine Comprehensive Cancer Center, Chicago, IL, USA.}
\affil[6]{\small Department of Physics and Astronomy, University of Iowa, Iowa City, Iowa 52242, USA.}
\affil[7]{\small Department of Applied Physics, Eindhoven University of Technology, Eindhoven, 5600 MB, The Netherlands.}
\affil[8]{\small CZ Biohub Chicago, LLC, Chicago, IL 60642, USA.} 
\affil[9]{\small Center for Molecular Engineering and Materials Science Division, Argonne National Laboratory, Lemont, IL.} 
\affil[ ]{\text {\textsuperscript{\textdagger} denis-candido@uiowa.edu, aesserkahn@uchicago.edu, and pmaurer@uchicago.edu}}

\date{}

\begin{document}

\maketitle 
\begin{abstract}
Nitrogen-vacancy (NV) based quantum sensors hold great potential for real-time single-cell sensing with far-reaching applications in fundamental biology and medical diagnostics. 
Although highly sensitive, the mapping of quantum measurements onto cellular physiological states has remained an exceptional challenge. 
Here we introduce a novel quantum sensing modality capable of detecting changes in cellular activity. 
Our approach is based on the detection of environment-induced charge depletion within an individual particle that, owing to a previously unaccounted transverse dipole term, induces systematic shifts in the zero-field splitting (ZFS). Importantly, these charge-induced shifts serve as a reliable indicator for lipopolysaccharide (LPS)-mediated inflammatory response in macrophages. Furthermore, we demonstrate that surface modification of our diamond nanoprobes effectively suppresses these environment-induced ZFS shifts, providing an important tool for differentiating electrostatic shifts caused by the environment from other unrelated effects, such as temperature variations. Notably, this surface modification also leads to significant reductions in particle-induced toxicity and inflammation. 
Our findings shed light on systematic drifts and sensitivity limits of NV spectroscopy in a biological environment with ramification on the critical discussion surrounding single-cell thermogenesis. Notably, this work establishes the foundation for a novel sensing modality capable of probing complex cellular processes through straightforward physical measurements.   
\end{abstract}

NV centers in diamond nanocrystals enable the probing of physical properties underlying complex biological processes. The remarkable sensitivity of NV centers to magnetic fields has enabled groundbreaking applications, including the detection of subcellular assemblies in magnetotactic bacteria\cite{Le_Sage2013-yv}, action potentials in individual neurons\cite{Barry2016-he}, and mitochondrial activities\cite{Nie2021-ls}. Similarly, NV-based nanothermometry has facilitated the control of thermal gradients in living organisms, offering insights into embryogenesis\cite{Choi2020-qg} and introducing novel approaches to neural stimulation\cite{Ermakova2017-vg}. Although promising, these current sensing modalities provide only limited information on cellular activity. Expanding the scope to monitor cellular activities through their chemical environment would unlock new opportunities, offering complementary insights beyond those accessible through magnetic field and temperature sensing. Such an approach could illuminate processes ranging from metabolic activities and cellular stress responses to progression through the cell cycle, apoptosis, differentiation pathways, and immune cell activation.

Our approach to probing cellular activities leverages the sensitivity of NV centers in diamond nanoparticles to environment-induced charge transfer. Each nanoparticle, hosting an ensemble of approximately 100 NV centers (Fig. \ref{fig:1}(a)), acts as a nanoscale sensor capable of detecting subtle changes in its surroundings. Recent findings\cite{Zvi2023-zd,Petrakova2012-wz,Xu2024-ig,Broadway2018-vw} have demonstrated that surface potential-induced band bending in diamond leads to the ionization of substitutional nitrogen defects (P1 centers), creating a localized charge layer. This charge layer, in turn, generates an electric field that can be precisely measured using NV-based optically detected magnetic resonance (ODMR) spectroscopy.
  
In this work, we establish a readout approach designed to probe environment-induced electric field changes in diamond nanocrystals through measurable shifts in their ODMR spectra away from their initial state. This measurement approach is based on a comprehensive model accounting for the influence of the commonly neglected secondary transverse dipole term ($d'$). We demonstrate the significance of our model in saline buffer and show that the observed ZFS shifts are effectively suppressed in diamond nanocrystals encapsulated in biocompatible core-shell structures --- further supporting our hypothesis that these effects result from surface interactions with the environment. Utilizing our newly developed sensing modality, we show that the identified ZFS shifts enable us to differentiate cellular activation states, as exemplified in RAW cells stimulated with LPS. Finally, we discuss the implications of the observed ZFS shift for ODMR-based sensing, particularly in nanoscale thermometry, and highlight its potential applications in monitoring a broad spectrum of cellular processes.

\subsubsection*{Sensing with randomly oriented spins} 

Conventional NV-based electric field sensing relies on coupling fields to the NV's axial $d_\parallel$ or transverse $d_\perp$ dipole moments\cite{Dolde2011-bu}. Crucially, these interactions depend on the relative orientation of each NV center with respect to the local electric field. In general, this orientation dependence results in an inhomogeneous broadening of the ODMR line\cite{Michl2019-rt} but not a static overall ZFS shift. The absence of any systemic shift makes it challenging to separate the effect of electric field-induced line broadening from other noise sources, such as magnetic fields. 

\begin{figure} [h!]
    \centering
    \includegraphics[width=0.5\linewidth]{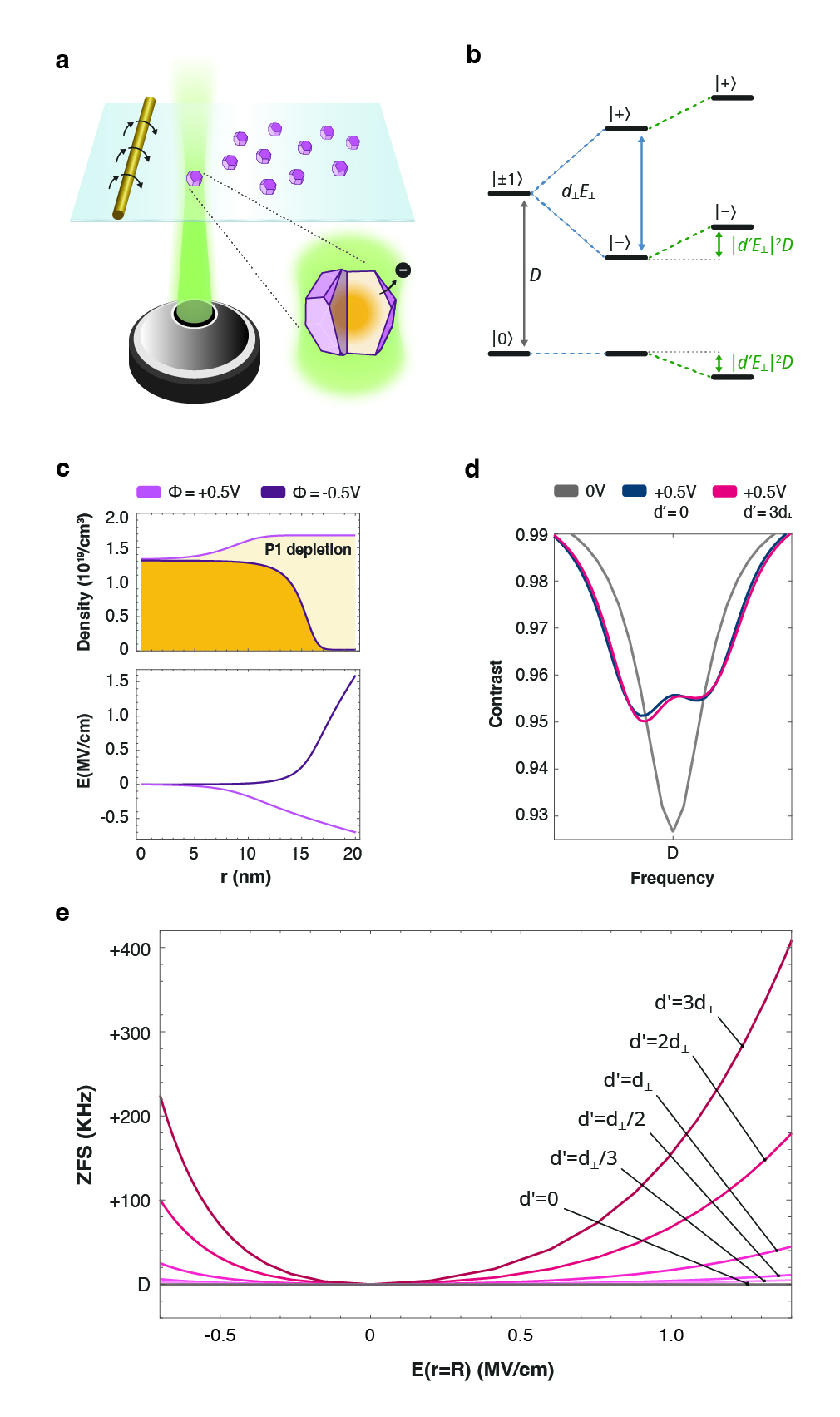}
    \caption{Theoretical framework for sensing electric fields with randomly oriented spins. (a) Illustration of the system's components, including diamond nanocrystals hosting randomly oriented NV ensembles, a Rabi drive, and a laser excitation that result in electron transfer between the diamond and the environment. The depletion of P1 centers from the surface inwards is illustrated by an orange gradient in the nanocrystal's slice. (b) NV energy level diagram illustrating both the effect of the dipole moment $d_{\perp}$ (blue arrow), and the second-order correction due to the dipole moment $d'$ (green arrows). (c) The bare diamond P1 density (upper panel) and the radial electric field profile (lower panel) shaped by a surface potential of $-0.5$~V (light purple) and $+0.5$~V. (d) Simulated ODMR curves composed of 100 randomly oriented NVs for no surface potential (gray), as well as for $\phi=-0.5$~V without ($d'=0$, black) and with ($d'=3d_{\perp}$, red) second order corrections. (e) The shift in ZFS as a function of electric field at the surface with $d'=0, \frac{d_{\perp}}{3}, \frac{d_{\perp}}{2}, d_{\perp}, 2d_{\perp}$, and $3d_{\perp}$.}
    \label{fig:1}
\end{figure}

In our approach, we go beyond this simple model and include the contribution from the transverse dipole term $d'$, which is estimated to be of similar magnitude as $d_{\perp}$ yet is typically neglected (Fig. \ref{fig:1}(b))\cite{Michl2019-rt,Policies-Practices2018-sa,Chen2020-we}. The full Hamiltonian then takes the form\cite{PhysRevB.110.024419,PhysRevB.85.205203}:

\begin{equation}
\frac{\mathcal{H}}{h} = (D + d_{\parallel}E_{z})(S_{z}^{2}-\frac{2}{3}) + d_{\perp}(E_{x}(S_{y}^{2} - S_{x}^{2}) + E_{y}\{S_{x}, S_{y}\}) + d'(E_{x}\{S_{x},S_{z}\} + E_{y}\{S_{y},S_{z}\}),
\end{equation}
where $\mathbf{E}=(E_x,E_y,E_z)$ denotes the electric field at the location of the NV center, $D$ the temperature-sensitive part of the ZFS, $h$ the Planck's constant, $\{\cdot,\cdot\}$ the anticommutator, and $\mathbf{S}=(S_x,S_y,S_z)$ the spin-1 matrices. 
Using second-order perturbation theory and assuming that the ensemble average satisfies $\left\langle E_z \right\rangle \to 0$, we find that the ODMR frequency shift is determined by
$\left\langle |E_\perp|^2 \right\rangle |d'|^2 / D$,
where $E_\perp = \sqrt{E_x^2 + E_y^2}$ represents the transverse field. 

We connect the local electric field to the surface potential ($\phi$) and the specific diamond doping profile. Following the procedure described in ref.\cite{Zvi2023-zd}, we solve Poisson's equation, accounting for the P1, vacancies, and NVs densities\cite{Deak2014-hw,Shenderova2019-lm} (Fig. \ref{figS: model}). Figure \ref{fig:1}(c) describes the P1 densities (upper panel) and the resulting electric field profiles (lower panel) for an idealized spherical particle, where $\phi=0.5$~V, accounts for the surface-potential of a bare oxygen-terminated diamond\cite{Zvi2023-zd}, and $\phi=-0.5$~V, the case of a nanocrystal that has undergone a depletion of P1 centers due to an electron transfer from the diamond to the environment. 

Finally, using Lindblad formalism (see methods and SI note \ref{SI theoretical model}), we simulate the ODMR spectrum for an average of 100, randomly oriented NV centers under continuous-wave laser excitation and microwave driving. Our simulations reveal several key properties of this approach. First, the presence of a surface potential $\phi=0.5$~V vs. $\phi=0$~V results in a splitting of the resonance into two eigenstates (Fig. \ref{fig:1}(d)). Second, the ensemble-average of the dipole terms results in asymmetric line broadening and contrast changes to the $f^+$ and $f^-$ transitions (the transitions from $\ket{0} \rightarrow \ket{+}$ and $\ket{0} \rightarrow \ket{-}$, respectively) that are dependent on the relative orientation of each NV to the localized electric field and to the Rabi drive field (SI note \ref{SI theoretical model}). Third, as predicted by our analytical result from eqn.~\ref{freqODMR}, the addition of $d'$ leads to an overall shift in ZFS. A systematic investigation of the ZFS resonance as a function of electric field reveals a quadratic field dependence on $d'$ (Fig. \ref{fig:1}(e)).

\begin{figure}[h!]
    \includegraphics[width=1\linewidth]{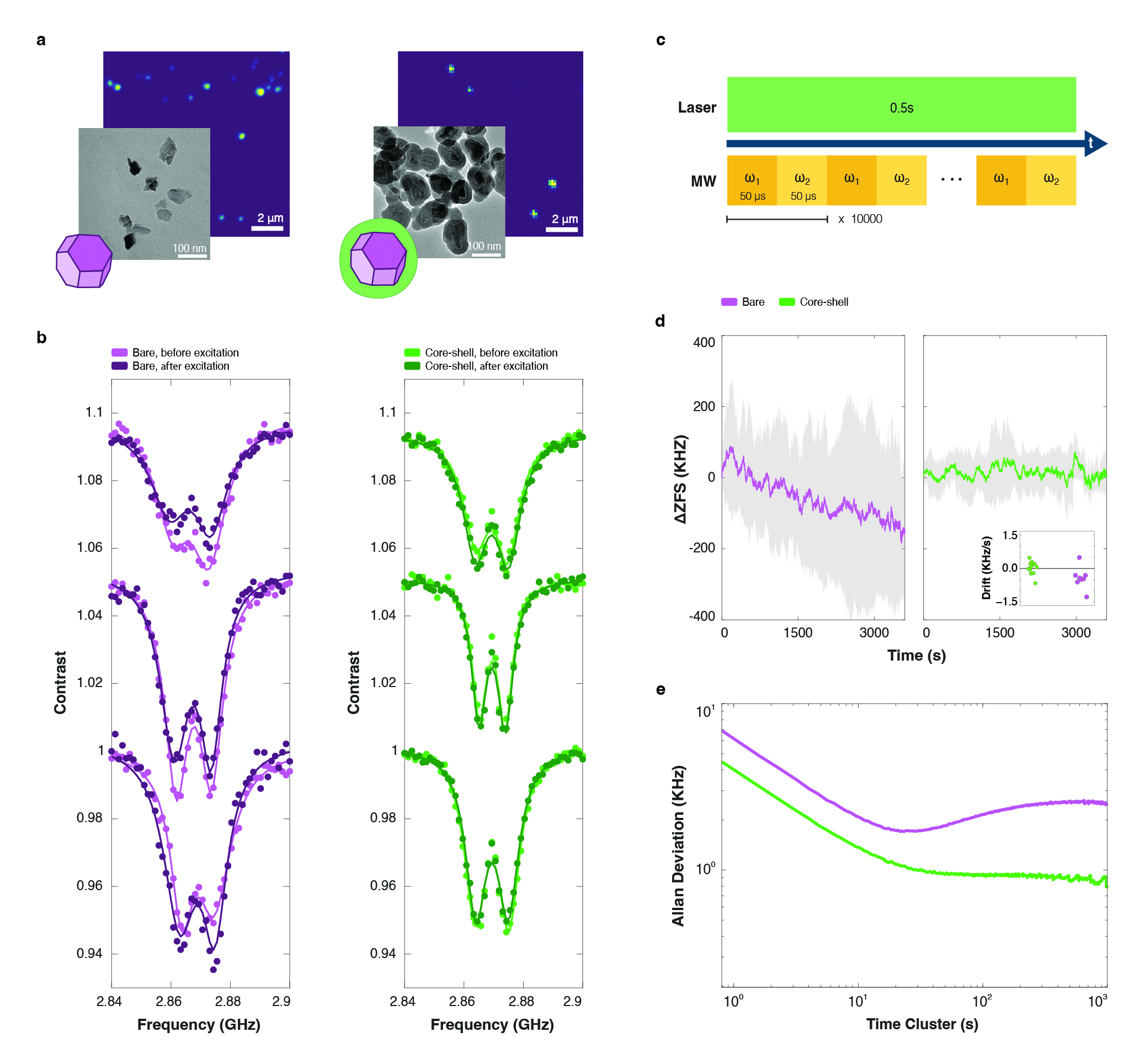}
    \caption{ZFS behavior in PBS. (a) Confocal and transmission electron microscopy (TEM) images of bare (left panel) and core-shell (right panel) particles. (b) Representative ODMR spectra for bare (left panel) and core shell (right panel) particles before (light color) and after (dark color) laser excitation. Points represent experimental data, while lines are double Lorentzian fits. (c) Pulse sequence used to track ZFS with higher temporal resolution. (d) Average ZFS time-series curves for bare ($N=8$, left panel) and core-shell ($N=12$, right panel) particles during laser excitation. One standard deviation regions are shown in shaded gray areas. Drift terms extracted are shown in the inset for both bare (purple points) and core-shell (green points) particles. (e) Allan deviation computed from weighted time series curves for bare (purple) and core shell (green) particles.}
    \label{fig:2}
\end{figure}

\subsubsection*{ZFS in phosphate-buffered saline environment}
Inspired by our theoretical findings, we investigate whether the properties predicted by our model are apparent in experiments under biologically relevant conditions. We deposited bare diamond nanocrystals with a diameter of 70 nm (Fig. \ref{fig:2}(a), left panel) on a coverslip and investigate the ZFS behavior in phosphate-buffered saline (PBS), which contains charged ions that mimic the physiological conditions of biological cells. Interestingly, $\sim$1 hour of laser excitation ($\sim\SI{0.1}{\milli\watt\per\micro\meter^{2}}$) led to significant alterations in the shape of the ODMR spectrum, including asymmetric broadening of the full-width half maximum (FWHM) and imbalanced changes in the contrast of the $f^+$ and $f^-$ transitions (Fig. \ref{figS: PBS} and Fig. \ref{fig:2}(b), left panel). These orientation-dependent changes are well predicted by our model (SI notes \ref{SI theoretical model} and \ref{SI ODMR fitting}). In contrast, the ZFS systematically shifted to lower frequencies by an average of $\SI{-0.45\pm 0.16}{\mega\hertz}$ after adding PBS. Such a decrease in the ZFS aligns with a charge transfer and a subsequent reduction in the intra-particle electric field sensed by the NV centers (Fig. \ref{fig:1}(e). See SI note \ref{SI ODMR fitting} for limitations in determining the magnitude of electric field changes). We note that we observed no systematic changes in ZFS when performing these measurements in air and water (Fig. \ref{figS: air and water}), suggesting that the observed ZFS shifts depend on the chemical environment.

Measuring the ZFS by acquiring a full ODMR spectrum is a time-consuming process that limits the ability to observe rapid ZFS changes. To capture ZFS dynamics with better temporal resolution, we implemented a two-point measurement scheme previously developed for NV thermometry\cite{Singam_2020}. In this scheme, we measure the fluorescence at two microwave frequencies (Fig. \ref{fig:2}(c)). An addition of a proportional-integral-derivative (PID) control loop and simultaneous spatial tracking (Fig. \ref{figS: rapid ZFS tracking}) enables a rapid and efficient extrapolation of the ZFS value (see Fig. \ref{figS: rapid ZFS tracking} and SI note \ref{SI Rapid ZFS tracking} for a summary of the measurement's performance in constant and modulated temperatures). We used our PID measurement approach to explore the rapid dynamics of the ZFS change in PBS under laser illumination and observed significant temporal drifts (Fig. \ref{fig:2}(d), left panel). In all observed cases we found that increased illumination time resulted in an larger decrease of the ZFS, suggesting that the observed drifts in ZFS are an optically induced effect.  

\subsubsection*{Reduced drifts in core-shell nanocrystals}
We hypothesized that exposure of the nanoparticles' surface to PBS is a key factor leading to the observed ZFS shifts. To test this hypothesis, we modified the surface by coating the particles with a \SI{15}{\nano\meter} thick protective silica shell\cite{Zvi2023-zd}. In contrast to bare diamond nanocrystals in PBS, the ZFS of core-shell particles remains stable during extended laser excitation, implying a stable surface potential. In fact, we find that the initially measured ODMR spectra (Fig. \ref{fig:2}(b), light green curve) are qualitatively unchanged after continuous laser excitation (Fig. \ref{fig:2}(b), dark green curve. For a time series of the ZFS see the right panel in Fig. \ref{fig:2}(d)). We note that prior to performing the measurements, we applied a short laser pulse in order to initialize a charge equilibrium in the diamond particle (Fig. \ref{figS: photoionization} and SI note \ref{SI Evidence for photo-assisted charge transfer}). The ZFS drift profile for bare ($N=8$) and core-shell ($N=12$) particles was then quantitatively characterized from their time-traces using a drift term analysis (methods), revealing negative drift terms for bare particles that significantly differ (see Extended Data Table \ref{extended_data_table1} for p-values) from the near zero drift terms of core shell particles in PBS (Fig. \ref{fig:2}(d), inset). For additional insight, we employed an Augmented Dickey-Fuller (ADF) test (See Methods' Statistics section) --- a statistical tool to test for non-stationary processes\cite{Dickey1979-ty}. A p-value of 0.557 indicates that the ZFS time series for bare nanocrystals  (Fig. \ref{fig:2}(d), left panel) is indeed a non-stationary process, whereas, for core-shell particles  (Fig. \ref{fig:2}(d), right panel) a p-value of \num{3.06e-4} suggests a stationary process. We further examined the stability of the ZFS by computing the Allan deviation for the time-domain signal (Fig.~\ref{fig:2}(e)). This suggests that for short-time averages ($\tau < \SI{10}{\second}$) the uncertainty of the ZFS is three times smaller in core-shell particles compared to their bare counterparts. Furthermore, the Allan deviation for bare particles reveals a systematic drift-term in the ZFS for times exceeding $\SI{20}{\second}$, whereas for core-shell particles the Allan deviation plateaus at $\SI{0.847}{\kilo\hertz}$, suggesting that the uncertainty in ZFS has reached a white noise limit.

\subsubsection*{Particle surface: Effects on toxicity and inflammation}
\begin{figure} [h!]
    \includegraphics[width=1\linewidth]{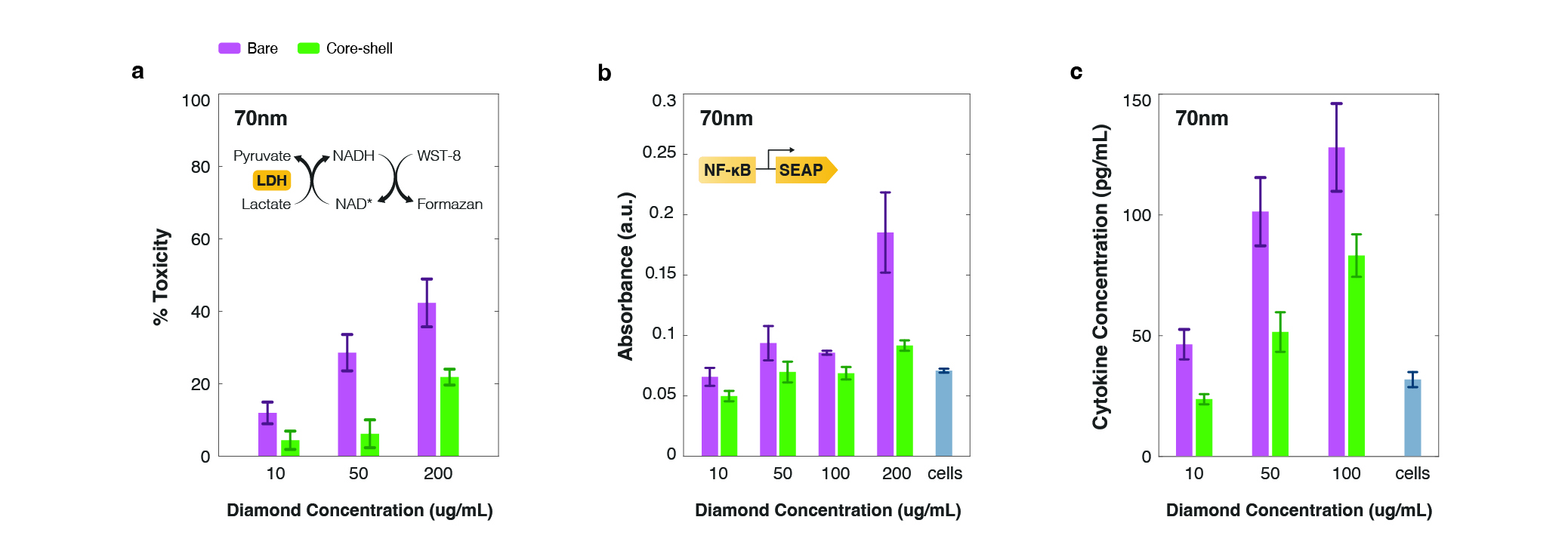}
    \caption{Toxicity and inflammation. (a) $\%$ toxicity measured with an LDH secretion assay for cells incubated  with different diamond core concentrations of 70~nm bare (purple) and core-shell (green) particles. (b) NF-$\kappa$B expression measured via Quanti-Blue assay for cells incubated  with different diamond core concentrations of 70~nm bare (purple) and core-shell (green) particles. (c) TNF-$\alpha$ secretion measured with a LEGENDplex assay for cells incubated  with different diamond core concentrations of 70~nm bare (purple) and core-shell (green) particles. Bright blue groups show measurement for cells with no particles. See methods and SI note \ref{SI Toxicity and Inflammation} for more information about the protocol. All p-values are detailed in Extended Data Table \ref{extended_data_table1}.}
    \label{fig:3}
\end{figure}

Recognizing the correlation between surface potential, cellular toxicity, and inflammation\cite{Moser2017-st,Seong2004-hb,Farrera2015-kt,Gustafson2015-en}, we evaluated the influence of bare and core-shell diamond nanocrystals on live cells. Previous reports of diamond nanocrystals toxicity in immune cells\cite{Chu2014-ng,Xing2011-ft,Thomas2012-kl,Zhang2010-wx} led us to focus on the particles' effect on RAW cells, a murine macrophage cell line. In our investigation we incubated diamond nanoparticles of 70~nm diameter at concentrations ranging from 10 to \SI{200}{\micro\gram\per{\milli\liter}} with approximately 12,000 RAW cells per well. To quantify toxicity, the supernatant was analyzed for lactate dehydrogenase (LDH) release following 6, 24, and 48 hours of incubation. Across all tested concentrations and time points, cells exposed to core-shell particles exhibited significantly lower LDH levels compared to those incubated with bare particles (Fig. \ref{figS: Toxicity and Inflammation}). Figure \ref{fig:3}(a) summarizes the percentage of cytotoxicity at the 24 hours time point.

We aimed to measure inflammation by monitoring nuclear factor-kappa B (NF-$\kappa$B) expression levels using a Quanti-Blue assay. Following overnight incubation, core-shell particles showed a significant reduction in inflammation at concentrations higher than \SI{50}{\micro\gram\per{\milli\liter}} (Fig. \ref{fig:3}(b)) when compared with bare particles. To better understand the inflammatory responses induced by diamond nanocrystals, we analyzed the cytokine release profile. Using a LEGENDplex assay, we measured 13 cytokines and chemokines and observed an inflammatory response solely characterized by elevated levels of TNF-$\alpha$ (Fig. \ref{fig:3}(c)). This pattern is consistent with inflammatory responses seen with other nanoparticles systems\cite{Moser2017-st,Khanna2015-eb}. Importantly, while both particle types induced higher levels of TNF-$\alpha$ secretion at concentrations above \SI{50}{\micro\gram\per{\milli\liter}} compared to control, cells incubated with core-shell particles exhibited significantly lower TNF-$\alpha$ levels compared to those exposed to bare particles (see Extended Data Table \ref{extended_data_table1}). We repeated the measurements with smaller, 40~nm diameter particles and found that core-shell particles again reduced the expression of LDH, NF-$\kappa$B, and TNF-$\alpha$ (Fig. \ref{figS: Toxicity and Inflammation} and SI note \ref{SI Toxicity and Inflammation}).

\subsubsection*{Detecting cellular inflammation through ZFS measurement}
\begin{figure}[h!]
    \includegraphics[width=1\linewidth]{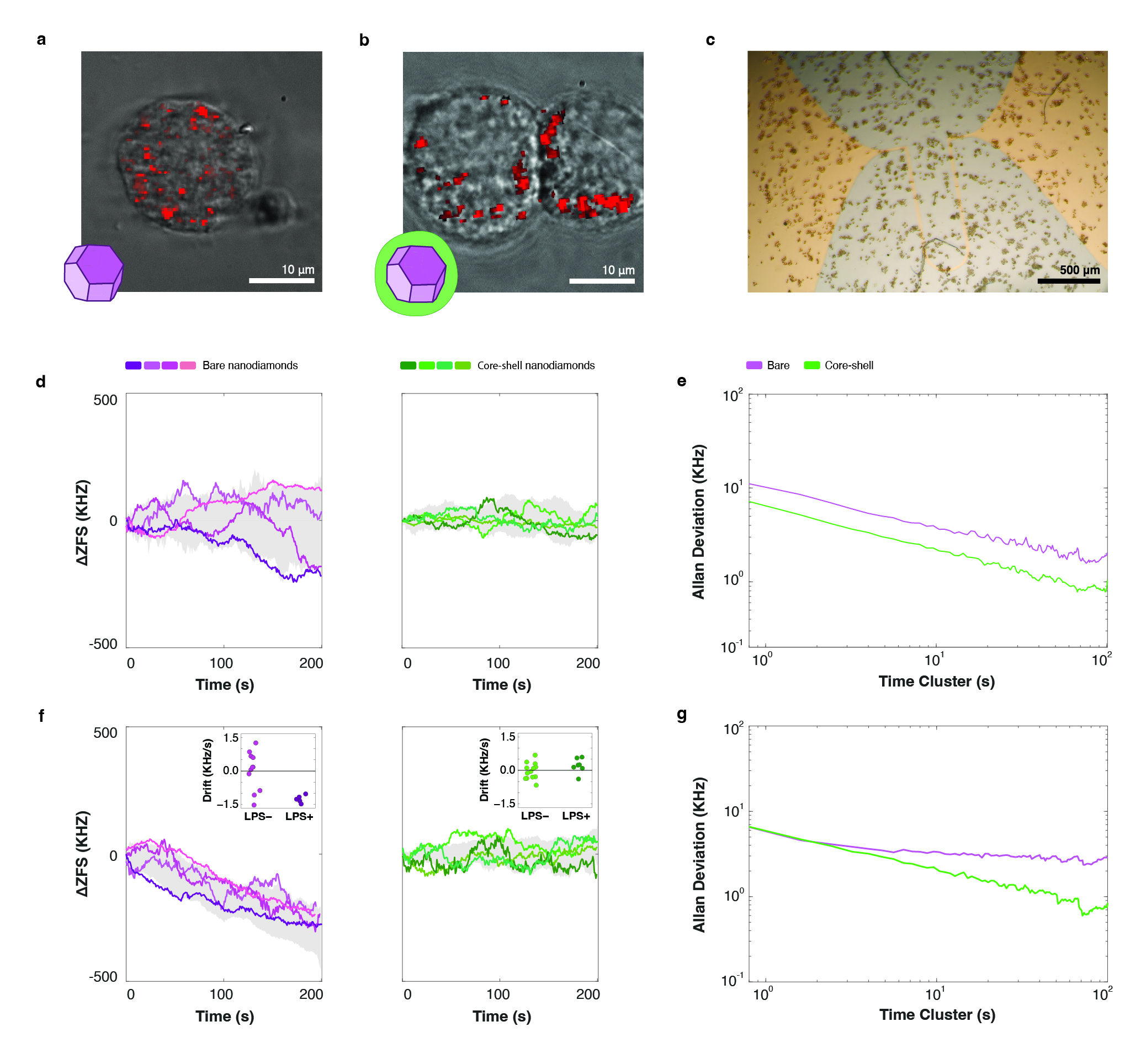}
    \caption{Probing macrophages inflammation with charge sensitive quantum nanoprobes. Cells were imaged following incubation with bare (a) and core-shell (b) particles. (c) Bright field image of cells growing in a customized well on top of an $\Omega$-shaped coplanar waveguide (SI note \ref{SI Coplanar waveguide}) for ZFS measurements in live cells. (d) ZFS time series curves of four representative bare (purple, left panel, total $N=11$) and core-shell (green, right panel, total $N=13$) particles incubated in nascent RAW cell, and their corresponding Allan deviations in (e). (f) ZFS time series curves of four representative bare (purple, left panel, total $N=5$) and core-shell (green, right panel, total $N=7$) particles incubated in LPS-stimulated RAW cell, and their corresponding Allan deviations in (g). Insets in (f) compare drift terms extracted for ZFS tracking of bare (left inset) and core-shell (right inset) particles in nascent and LPS-activated cells.}
    \label{fig:4}
\end{figure}

We conclude our study by demonstrating our sensor's ability to detect cellular activity. As a model system, we focus on the inflammatory response of macrophage-like RAW cells to the TLR-agonist LPS. We first incubated cells with bare (Fig. \ref{fig:4}(a)) or core-shell (Fig. \ref{fig:4}(b)) particles and investigated the ZFS time-series with no added stimulation (Fig. \ref{fig:4}(c)). The ZFS was tracked over 200 seconds under continuous laser excitation for 11 bare nanocrystals in 7 cells and 13 core-shell nanocrystals in 9 cells (Fig. \ref{figS: ZFS in RAW cells}). The short measurement time was chosen to minimize phototoxicity and preserve cellular viability. While the investigated diamond nanocrystals exhibited significantly larger ZFS fluctuations compared to those measured in PBS, we did not observe drifts in ZFS for either bare or core-shell particles (Fig. \ref{fig:4}(d)). Notably, bare diamond nanocrystals showed significantly wider variations in their ZFS values compared to core-shell particles. This trend was also evident in the Allan deviation analysis, which indicated lower variability and greater stability for core-shell particles (Fig. \ref{fig:4}(e)).

To assess ZFS behavior in inflamed cells, we stimulated RAW macrophages with \SI{1.5}{\micro\gram\per{\milli\liter}} of LPS following incubation with bare and core-shell diamond nanocrystals. Notably, all tested intracellular bare nanocrystals (5 particles in 4 cells) exhibited a systematic ZFS drift over a course of 200-second (Fig. \ref{fig:4}(f)), with an average total shift of $-0.27\pm\SI{0.03}{\mega\hertz}$. Such a shift would correspond to a temperature increase of \SI{3.62}{\degreeCelsius} over the course of the measurement. To rule out the possibility that observed drifts are a result of LPS-induced temperature increase, we repeated the measurement on cells incubated with core-shell particles (7 particles in 5 cells). These measurements showed no sizable ZFS drifts, in strong agreement with our thermodynamical analysis, suggesting that the chemical environment and not temperature is the cause for the observed effects in bare particles (SI note \ref{SI Thermodynamic analysis of RAW cell thermogenesis}; for temperature sensitivity of bare and core-shell particles in cells see Fig. \ref{figS: ZFS in RAW cells} and SI note \ref{SI ZFS tracking during temperature modulation}). Note, for experimental reasons ZFS tracking was initiated at various time points, ranging from $t=20$ to $t=100$ minutes post-stimulation, and in all cases continued for 200 seconds per measurement to compare the dynamics with that seen in nascent cells. A more careful assessment reveals that all bare diamond nanocrystals in LPS-activated cells exhibited a negative ZFS drift term (see methods) that significantly differ (t-test p value of \num{6.38e-03}) from those obtained for nascent cells (Fig. \ref{fig:4}(f), left inset). On the other hand, core-shell particles showed no notable differences compared to nascent cells (Fig. \ref{fig:4}(f), right inset). Interestingly, the extracted drift terms in cells were about an order of magnitude larger than those observed in PBS (Fig. \ref{fig:2}(d), inset), indicating a faster shift in electric field. An Allan deviation analysis demonstrates the influence of the drift on measurement sensitivity, with bare particles exhibiting significantly higher deviation at longer cluster times (Fig. \ref{fig:4}(g)).

\subsubsection*{Discussion and outlook}
We propose that the ZFS shifts reported here are a result of surface-potential-induced band bending that changes the charge states of P1 centers in the lattice. Theoretical\cite{Deak2014-hw} and experimental\cite{Petrakova2012-wz,Xu2024-ig,Broadway2018-vw} studies complement our recent electron paramagnetic resonance (EPR) findings in core-shell particles\cite{Zvi2023-zd} and strongly support the susceptibility of P1 centers to surface potential changes. The existence of a charge transfer process is further supported by the increase in ZFS shifts with increased laser power (Fig. \ref{figS: photoionization} and SI \ref{SI Evidence for photo-assisted charge transfer}), pointing towards the well-documented photo-assisted ionization of P1 centers\cite{Xu2024-ig,Policies-Practices2019-yh}. A steady decrease in PL over the measurements in PBS (Fig. \ref{figS: PBS}) suggests that this photoionization might also affect NV centers, while the enhanced PL stability of core-shell particles points to the suppression of charge transfer between the diamond and its environment. In fact, recent results have shown that such charge transfer can be suppressed in diamond with a dielectric passivation layer\cite{Zvi2023-zd,Kumar2024-ix}, as evident in this work by the suppression of ZFS shifts in core-shell particles.  

While the exact mechanism leading to the observed shifts in PBS and stimulated RAW cells merits further investigation, the ZFS shifts towards lower frequencies indicate an overall reduction in the electric field experienced by the NVs. The mitigation of charge-transfer-mediated shifts in core-shell particles allows us to postulate parts of the process. One possibility is that in PBS, ions in the solution may interact with the bare diamond surface, affecting surface potential and charge exchange rates. Additionally, components in the saline solution may change the redox potential near the diamond surface, leading to favorable electron exchange. A depletion of donor P1 centers near the silica-diamond interface\cite{Zvi2023-zd}, as well as the energy barrier provided by the silica shell, should significantly slow further electron transfer from the ND interior to the environment. Additionally, we recently showed that the silica shell reduces the density of surface defects and other spins\cite{Zvi2023-zd}, which are commonly associated with electron and hole trapping. This trapping phenomenon, widely documented for bulk diamond\cite{Sangtawesin2019-jb,Stacey2019-xi} and bare nanoparticles\cite{Bian2021-xs}, is greatly suppressed in core-shell structures. Together, these mechanisms provide evidence for a surface-induced effect rather than temperature, highlighting our core-shell particles as an efficient method for decoupling the two effects, thus providing a useful tool for enhanced thermometry in complex biological settings.
 
In nascent RAW cells, we attribute fluctuations in the ZFS to yet-unidentified processes in the complex intracellular environment. The lack of a systematic drift in bare particles is aligned with cellular homeostasis balancing biophysical parameters like oxidative and reducing processes\cite{Muri2021-vc,Morris2022-sn} or pH in non-stimulated cells. Upon LPS stimulation, these immune cells are expected to experience a Toll-like-receptor (TLR)-4 dependent inflammatory response that leads to M1 polarization\cite{Zhu2015-xi} and an associated metabolic reprogramming\cite{Liu2021-my,Kelly2015-en}. One possibility is that the shift towards a more oxidative environment\cite{Muri2021-vc,Morris2022-sn} decreases the negative electric field in the particle by ionization of P1 donors, which in turn leads to the experimentally observed reduced ZFS. Nonetheless, identifying the exact metabolic process and its precise interaction with our diamond sensor is an outstanding question. Decoupling direct reactions with the sensors' surface groups, nanoscale pH changes, and electrochemical potential changes in the environment would require experiments with minimal models to isolate specific influences, an effort we are currently pursuing. 

Another interesting direction stems from the reduced toxicity and inflammation of core-shell compared to bare particles. The reduced toxicity of core-shell particles is in good agreement with our previous findings based on other types of silica-coated particles\cite{Moser2017-st}, as well as other reports using silica coated diamond nanocrystals\cite{Chu2014-ng,Rehor2014-ek,Bumb2013-dd,Prabhakar2013-vu}. As a possible mechanism explaining the effects on toxicity, we propose that the surface potential of particles and their interaction with biomolecules play a central role\cite{Seong2004-hb,Farrera2015-kt,Gustafson2015-en}. As such, the photo-induced charge exchange between the NDs and their surroundings could exacerbate cytotoxic effects, and thus merits further investigation. 

Our findings also have ramifications for ODMR-based sensing applications. Our model predicts the asymmetric line broadening and PL contrast of the $f_{\ket{0}\rightarrow\ket{+}}$ and $f_{\ket{0}\rightarrow\ket{-}}$ transitions (Fig. \ref{fig:1}(d), Fig. \ref{figS: model}, and SI note \ref{SI ODMR fitting}), which have striking implications on the sensitivity limitations of ZFS measurements in cells\cite{Kucsko2013-fn,Choi2020-qg,Fujiwara2020-el,Petrini2022-cb}. More importantly, by considering $d'$, we provide a thermodynamically consistent, novel interpretation for the heavily debated systematic ZFS shifts reported in cellular systems\cite{Fujiwara2020-el,Petrini2022-cb, Baffou2014-rb,Sotoma2021-mm} that have typically been considered to result from local temperature changes\cite{Acosta2010-sd}.

Ultimately, our study introduces a novel sensing modality that links measurable ZFS shifts to changes in cellular activity. The new understanding of the interplay between the cellular environment and the particle's internal charge profile presents an opportunity for tailoring sensing strategies to monitor complex cellular processes including cellular differentiation, apoptosis, carcinogenesis, and other metabolic reprogramming events through a simple measurement with a quantum sensors. We are currently pursuing such strategies for distinguishing macrophage polarization states, as well as antigen specific T-cell induced apoptosis of tumor cells.

\subsection*{Materials and Methods}
\label{Materials and Methods}

\subsubsection*{Theoretical model from first principles}
\label{Methods Theoretical model from first principles}
Our theoretical ODMR results are simulated using a Lindblad density matrix formalism accounting for the laser excitation, microwave field driving the Rabi oscillations, and all radiative and non-radiative decays. We have also included relaxation and dephasing processes in the ground state of the NV center. In our formalism, all these processes are described by the Lindblad operators $L_k=\sqrt{\Gamma_{ab}}\ket{a}\bra{b}$ with rate $\Gamma_{ab}$ associated to the $\ket{b}\rightarrow\ket{a}$ transition. In the interaction picture, our Lindblad equation for the NV-$j$ reads 

\begin{equation}
    \frac{\partial\rho_j\left(t\right)}{\partial t}=\frac{i}{\hbar}\left[\rho_j\left(t\right),H_j\right]+\sum_{k}\left[L_{k}\rho_j\left(t\right)L_{k}^{\dagger}-\frac{1}{2}\left\{ \rho_j\left(t\right),L_{k}^{\dagger}L_{k}\right\} \right],
\end{equation}

with $\{A,B\}=AB-BA$ and $H_j$ describing the Hamiltonian for the NV-$j$. The total photoluminescence of all NVs is obtained via ${\textrm{PL}(t)=\sum_{j}\textrm{Tr}_{\textrm{ES}_j}[\rho_j(t)]}$ where the trace is taken with respect to the excited states (ES$_j$) of the NVs. The Hamiltonian $H_j$ depends on the electric field via Eq.~(\ref{hamiltonian}), which is calculated for a spherical ND via Poisson's equation 

\begin{equation}
    \nabla^2 \phi(\mathbf{r}) = -\rho(\mathbf{r})/\varepsilon,
\end{equation}
with electrostatic potential $\phi(\mathbf{r})$, charge density $\rho(\mathbf{r})$ and diamond dielectric constant $\varepsilon$. The Poisson equation is numerically solved in a self-consistent way with charge density accounting for the density of Nitrogen, vacancy,  NV centers, conduction, and valance bands. We solve the equation above with boundary conditions $\phi(r=R)=\phi_S$ and $\frac{d\phi(r=0)}{dr}=0$, with surface electrostatic potential $\phi_S$.

For the representative P1 center density, electric field profile, and ODMR curves in Fig. \ref{fig:2}, $\phi = -0.5$~V was calculated using a bare particle with a band-bending profile described in ref. \cite{Zvi2023-zd}. However, modeling the non-spherical, disc-like shape expected for our diamond nanocrystals\cite{Eldemrdash2023-wx} as a sphere would lead to underestimation of the electric field experienced in deeper NVs. Unfortunately, solving the Poisson's equation for a disc-like shape is non-trivial. Therefore, an effective diameter of $40$~nm was chosen for simulating the reported $70$~nm particles in this work. 

\subsubsection*{Diamond nanocrystals}
\label{Methods Diamond nanocrystals}
70 nm diamond nanocrystals were obtained from Adámas Nanotechnologies Inc. In brief, type 1b microcrystals are manufactured by static high-pressure, high-temperature (HPHT) synthesis and contain about 100-200 ppm of substitutional N. These particles are milled, irradiated with 2-\SI{3}{\mega\eV}, and annealed at 850 °C for 2 hrs by Adámas Nanotechnologies Inc\cite{Shenderova2019-lm}. 

\subsubsection*{Synthesis of core-shell particles}
\label{Methods Synthesis of core-shell particles}
The growth of uniform Silica shells on diamond nanocrystals was performed using a modified Polyvinylpyrrolidone (PVP) based technique\cite{Graf2003-ai,Rehor2014-ek} that is described in detail in ref. \cite{Zvi2023-zd}. See SI note \ref{SI Core-Shell synthesize} for a description of exact modifications to coat 70~nm particles.

\subsubsection*{TEM Characterization} 
\label{Methods TEM Characterization}
Bare and core-shell diamond nanocrystals were deposited on a copper (with a formvar carbon film) or Silicon (Si3N4 film) grid while ensuring minimal aggregation, as described in ref. \cite{Rehor2014-sq}. Images were taken using an FEI Tecnai G2 F30 300kV TEM.

\subsubsection*{RAW-Blue cells}
\label{Methods RAW-Blue cells}
RAW-Blue™ Cells(InvivoGen), an NF-$\kappa$B-SEAP reporter Cell line derived from the murine RAW 264.7 macrophages, were cultured (SI note \ref{SI RAW cells}) in T75 flasks (Thermo Fisher). For testing, cells were harvested by scraping and resuspended in 96-well plates for toxicity and inflammation assays, or in a custom-made printed circuit board (PCB) design with a coverslip customized for ZFS measurements (Fig. \ref{fig:4}(c) and SI note \ref{SI Coplanar waveguide}). Cells were allowed to adhere for 6-12 hours before incubation with bare or core-shell particles.

\subsubsection*{Measuring ZFS}
\label{Methods Measuring ZFS}
ODMR spectra were taken using a standard continuous-wave ODMR sequence with \SI{100}{\micro\sec} microwave (MW) pulse duration at an output power of 10-20 dBm from the amplifier. The ZFS was defined as $\frac{f^+ + f^-}{2}$, where $f^+$ and $f^-$ were obtained from a double Lorentzian fit (SI note \ref{SI ODMR fitting}). Rapid ZFS measurements were done using an adapted 2-point measurement\cite{Singam_2020}, with an added PID controller and a particle tracking algorithm (Fig. \ref{figS: rapid ZFS tracking} and SI note \ref{SI Rapid ZFS tracking}). In brief, to estimate the ZFS of the NV center over time, given a starting center frequency $\omega_c$, we measure the fluorescence at two microwave frequencies, $\omega_1 = \omega_c - \frac{\gamma}{2}$ and $\omega_2 = \omega_c + \frac{\gamma}{2}$, where $\gamma$ is the full-width half maximum of the ODMR spectrum of the NV. The fluorescence values at these frequencies, $I_1$ and $I_2$, are in the quasi-linear regime of the ODMR Lorentzian spectrum so that changes in their values correspond linearly with changes in the ZFS. As such, measuring $I_1$ and $I_2$ repeatedly allows us to form updated predictions for the new ZFS. We implement sideband modulation to enable rapid switching of the microwave output between $\omega_1$ and $\omega_2$ and feed the predicted ZFS value through a proportional-integral-derivative (PID) control loop to mitigate noise and fluctuations in the readings (SI note \ref{SI Rapid ZFS tracking}). Running this process continuously over our measurement, we generate estimated ZFS values over time.

\subsubsection*{ZFS measurements in air, water, and PBS}
\label{Methods ZFS in air,water, and PBS}
A drop of NDs (\SI{50}{\micro\gram\per\milli\liter} in water) was deposited on a plasma treated No. 1.5 glass coverslip and shaken for 10 minutes at 100 rpm in a humid chamber to prevent evaporation. After water removal, the coverslip was mounted on a custom PCB, and a coated \SI{25}{\micro\meter} wire was connected across as an RF antenna. A channel (Ibidi bottomless 6 channel slide) was attached to allow delivery and exchange of liquid using a syringe pump (InfusionONE) while the sample was mounted on the microscope stage. ODMR spectra and $\sim$60-minutes ZFS tracking data were obtained for bare and core-shell particles in air ($N_{bare}=8$ and $N_{core-shell}=8$), water ($N_{bare}=7$ and $N_{core-shell}=4$), and PBS ($N_{bare}=9$ and $N_{core-shell}=12$). 

\subsubsection*{Toxicity and inflammation assays}
\label{Toxicity and inflammation assays}
For all toxicity and inflammation assays, RAW-Blue cells were seeded in 96-well plates at 12,000 cells / well and incubated overnight at \SI{37}{\degreeCelsius} with 5\% of CO$_2$.

\subsubsection*{\textit{LDH assay}}
To quantify toxicity, CyQUANT LDH Cytotoxicity Assay (Invitrogen Cat. Nos. C20300) was performed.  Cells were incubated in triplicates with 10, 50, and \SI{200}{\micro\gram\per{\milli\liter}} of bare or core-shell particles for 6, 24, and 48 hours, before \SI{50}{\micro\liter} of supernatant was transferred from each sample to flat-bottomed 96 well plates for colorimetric measurements (SI note \ref{SI Toxicity and Inflammation}).

\subsubsection*{\textit{NF-$\kappa$B assay}}
To quantify NF-$\kappa$B expression, a QUANTI-Blue™ assay (InvivoGen Cat. code rep-qbs) was performed. Cells were incubated in triplicates with 10, 50, 100, and \SI{200}{\micro\gram\per{\milli\liter}} of bare or core-shell particles for 16 hours before \SI{20}{\micro\liter} of supernatant was transferred from each sample to flat-bottomed 96 well plates for colorimetric measurements (SI note \ref{SI Toxicity and Inflammation}).

\subsubsection*{\textit{Cytokine expression assay}}
Cytokine expression was measured using a LEGENDplex$^TM$ Mouse Inflammation Panel (BioLegend 740150). Cells were incubated in triplicates with 10, 50, and \SI{100}{\micro\gram\per{\milli\liter}} of bare or core-shell particles for 16 hours before \SI{15}{\micro\liter} of supernatant was transferred from each sample to the assay-provided V-bottom 96 well plates for addition of Capture Beads, Detection Antibodies, and SA-PE, followed by flow cytometry (SI note \ref{SI Toxicity and Inflammation}).   

\subsubsection*{ZFS measurements in live cells}
Approximately 12,000 RAW-Blue cells (InvivoGen) were seeded in a well (Ibidi 18 well bottomless $\mu$-Slide) attached to a coverslip containing a fabricated co-planar waveguide (SI note \ref{SI Coplanar waveguide}). Cells were incubated in media with \SI{20}{\micro\gram\per\milli\liter} of bare or core-shell particles for $\sim$6 hours. After incubation, the cells were washed twice and clear media was added to facilitate imaging. Measurements were done using our custom-built microscope (SI note \ref{SI Imaging system}), equipped with a live cell imaging chamber (Invivo Scientific, STEV.ECU.HC5 STAGE TOP), and a temperature controller. Since the live-cell chamber has temperature stability of $\pm$\SI{0.5}{\degreeCelsius} for our sensitive measurement, we added a closed-loop resistive heater (HT19R), temperature transducer (AD590 in media and HT10KR1 for cells), and controller (TC300) from Thorlabs, which allowed us to achieve temperature stability of $<$\SI{0.1}{\degreeCelsius}. ODMR spectra and ZFS tracking data were collected from 1-2 NDs from each cell. Nascent cells were randomly chosen and NDs were measured immediately ($N=11$ for bare and $N=13$ for core-shell). For inflammation conditions ($N=5$ for bare and $N=7$ for core-shell), cells were stimulated with \SI{1.5}{\micro\gram\per\milli\liter} of LPS and measured at t=30-90 min post stimulation. In all cases, ZFS tracking was limited to 200 seconds. In addition, all experiments were halted at total time t<180 minutes and cells were monitored using wide-field to confirm viability.

\subsubsection*{Statistics}
\label{Methods statistics}

\subsubsection*{\textit{ADF test}} 
To assess the stationarity of the time series data, we performed the ADF test, which evaluates the presence of a unit root—a key indicator of non-stationarity. The test equation included a constant and lagged differences to account for autocorrelation, with the optimal lag length determined using the Akaike Information Criterion. A p-value below 0.05 was considered evidence to reject the null hypothesis, indicating that the series is stationary. See SI note \ref{SI statistics} for more detailed information.

\subsubsection*{\textit{Allan deviation}}
To compute Allan deviations for individual NV ZFS time-series data, we recorded the raw fluorescence data for $I_1$ and $I_2$ and artificially generated data sets for measurements with larger averaging time by summing over subsequent fluorescence data (details in SI note \ref{SI statistics}). To compute the average Allan variance plot for the time-series ZFS data in PBS of bare (n = 8) and core-shell (n = 12) nanodiamonds in PBS, we applied a weighted average of the individual Allan variance plots with the weights being the total measurement duration. This method ensures that the aggregate plot accurately reflects the contribution of each time series based on its respective measurement period.

\subsubsection*{\textit{Toxicity and inflammation bar plots}}
All error bars represent standard error unless otherwise noted. All analyses were done using two-tailed t-tests. See Extended data table \ref{extended_data_table1} for specific p-values.

\subsubsection*{\textit{Intracellular ZFS measurements}}
A time series analysis was performed in order to extract drift terms for ZFS tracking data in cells. We first obtained a distribution of differences between consecutive data points and extracted the relevant drift term by calculated the deviation of the mean from zero (Fig. \ref{figS: ZFS in RAW cells}). The significance between data obtained in nascent and LPS-activated cells was calculated using a two-tailed t-test.

\subsubsection*{Data availability} 
The data that support the finding of this study are available from the corresponding authors upon reasonable request.

\subsubsection*{Acknowledgments}
During the preparation of this manuscript, we became aware of an independent study prepared by Prof. Fedor Jelezko's group addressing questions related to ZFS drifts in cells. Their findings are complementary to ours, further supporting the presented conclusion. We thank Michelle Klosinski for extensive work on the illustrations in the manuscript We thank Prof. Abraham Wolcott, Prof. David Simpson, Prof. Shimon Kolkowitz, and Prof. Ania Jayich for insightful discussions related to charge dynamics in diamond nanoparticles, Prof. Alexander Pearson for discussion on medical applications, and Prof. Sihong Wang, Prof. Shrayesh Patel, and Sean Sutyak for insightful discussion related to electrochemistry. We also thank Lingjie Chen, Jacob Feder, Xiaofei Yu, and Evan Villafranca for support with instrument development and control. U. Z., A. E.-K., and P. C. M. acknowledge financial support through QuBBE QLCI (NSF OMA- 2121044). M. S., A. E.-K., and P. C. M. are CZ Biohub Investigator. This work made use of the Pritzker Nanofabrication Facility part of the Pritzker School of Molecular Engineering at the University of Chicago, which receives support from Soft and Hybrid Nanotechnology Experimental (SHyNE) Resource (NSF ECCS-2025633), a node of the National Science Foundation’s National Nanotechnology Coordinated Infrastructure. This work also made use of the shared facilities at the University of Chicago Materials Research Science and Engineering Center, supported by the National Science Foundation under award number DMR-2011854. We also acknowledge the assistance of Yimei Chen, Dr. Jotham Austin, and Dr. Tera Lavoie from The University of Chicago Advanced Electron Microscopy Core Facility (RRID:SCR-019198).

\subsubsection*{Authors contributions} 
U.Z., D.R.C, A.E.K., and P.M. conceived and designed the study. U.Z., S.M., and D.O. designed and performed the experiments. U.Z., S.M., D.O., A.R.J., and S.W. designed the experimental instruments and software control. D.R.C performed theoretical analysis. D.R.C and M.E.F. provided theoretical model guidance and discussion. U.Z., D.O., and Q.C. designed and performed cellular assays. M.R. guided the LEGENDplex analysis and M.F. contributed towards toxicity and LEGENDplex experiments. K.O., M.C.G, M.S., D.R.C., A.E.K, and P.M. provided supervision and guidance. U.Z., P.M., D.O., and S.M. wrote the manuscript with extensive input from all authors.

\subsubsection*{Competing interests}
The work covered in this manuscript is the subject of a patent application filed by the authors’ institutions in the US Patent and Trade Office. 

\clearpage
\bibliography{references}
\clearpage

\renewcommand{\thefigure}{E\arabic{figure}}
\renewcommand{\thetable}{E\arabic{table}}
\setcounter{figure}{0} 
\setcounter{section}{0}
\setcounter{table}{0}

\section*{Extended Data}

\begin{figure} [h!]
    \includegraphics[width=1\linewidth]{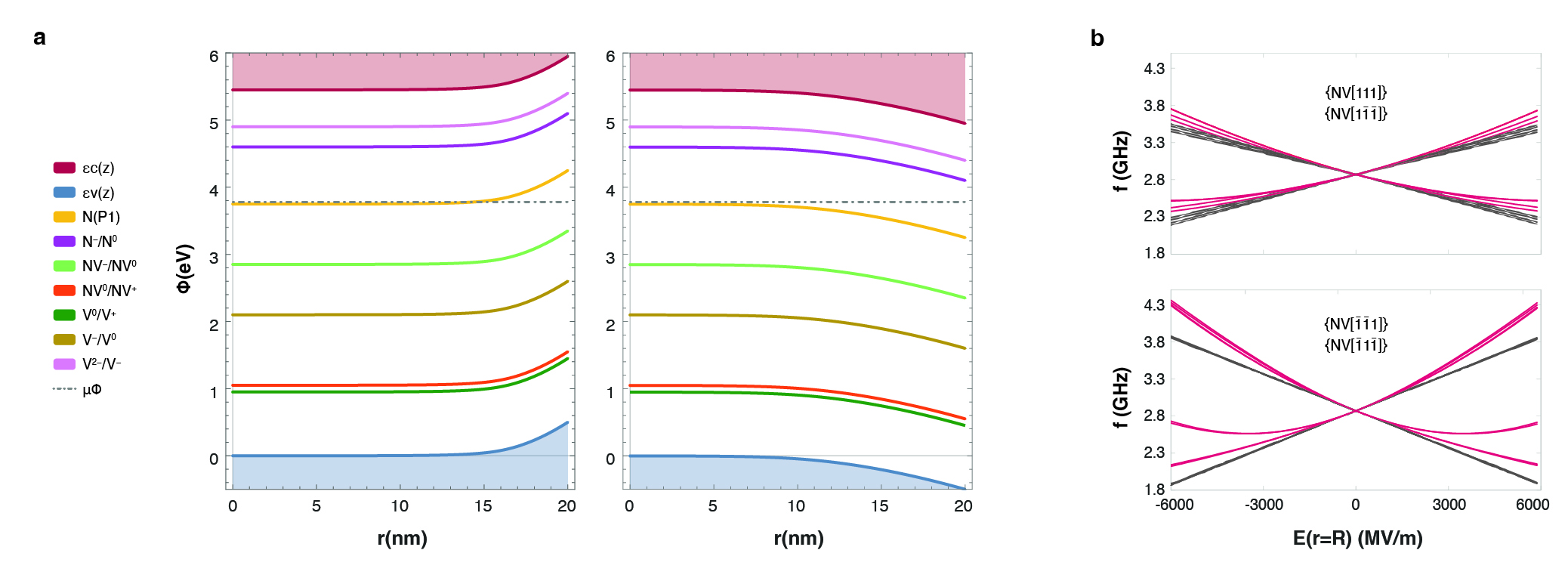}
    \caption{Theoretical model. (a)~Band bending simulation of upward (left panel) and downward (right panel) band bending with surface potential of $-0.5$V and $0.5$V, respectively. The downward band bending in carboxylated diamond nanocrystals is bending upwards as electrons transfer from the diamond to the environment. (b)~$f^-{\rightarrow}f^+$ transition frequency for different NV orientations as a function of surface (maximum) electric field for $d'=0$ (black lines) and $d'=d_{\perp}$ (red lines). The upper panel shows the orientations and anti-orientations of the [111] and the $[1\overline{1}\overline{1}]$ NVs. The lower panel shows the orientations and anti-orientations of the $[\overline{1}\overline{1}1]$ and the $[\overline{1}1\overline{1}]$ NVs. The unidirectional shift in ZFS is apparent for all orientations. Upon change in the chemical environment (and likely promoted by laser illumination), a charge transfer between the diamond and the environment will change the initial electric field experienced by the NVs. Consequently, a perturbation of the eigenstates of individual NVs leads to an overall shift of the ZFS towards higher or lower frequencies in response to an increase or decrease in electric field, respectively.}
    \label{figS: model}
\end{figure}
   
\clearpage
\begin{figure} [h!]
    \includegraphics[width=1\linewidth]{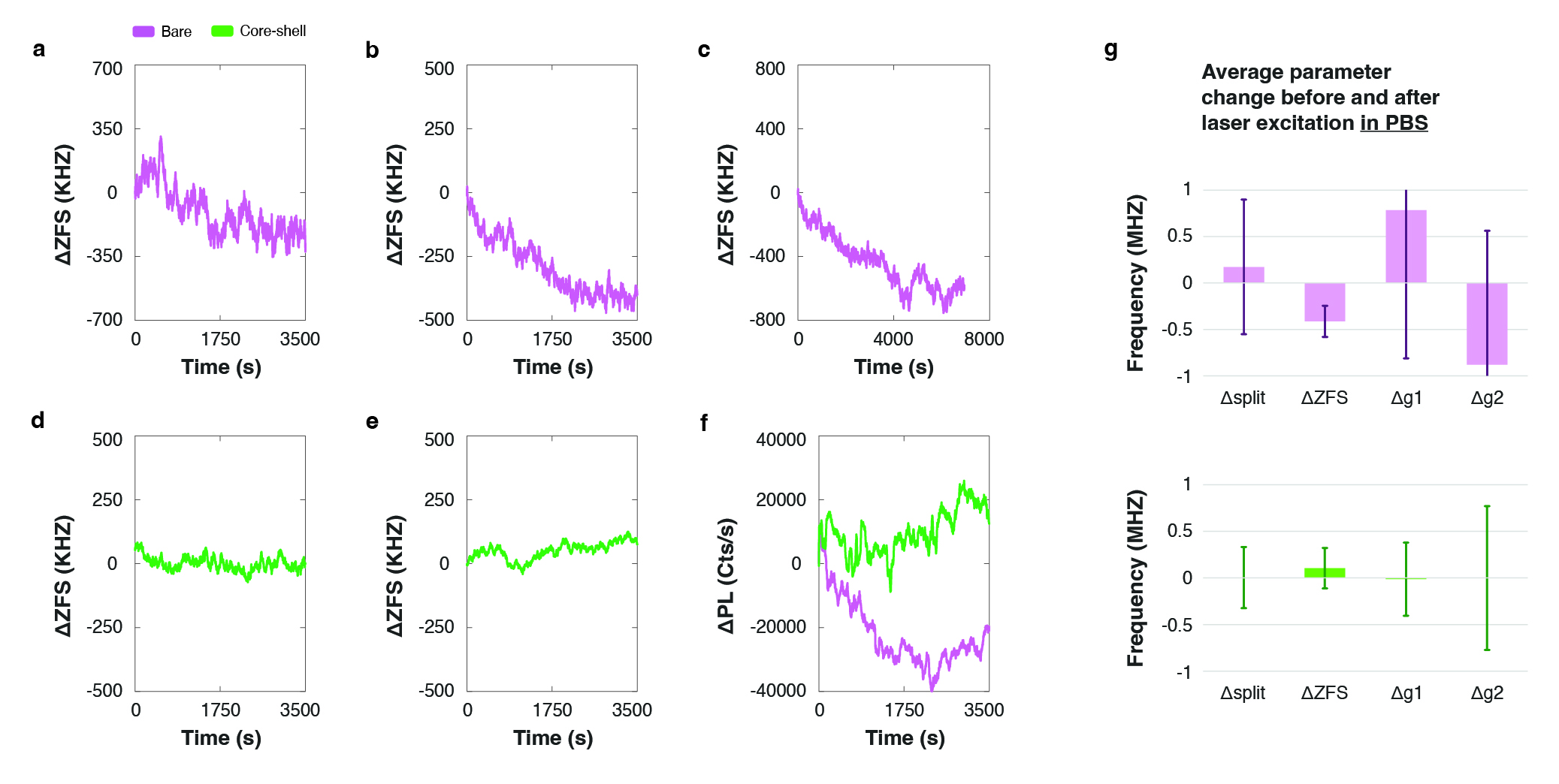}
    \caption{ZFS behavior in PBS. (a-b)~Representative ZFS tracking curves during $sim$1~hr of laser excitation of bare particles in PBS. $N=8$ data sets were averaged to produce the left panel of Fig. \ref{fig:2}(d). (c)~A representative longer time trace of the data showed in (b), demonstrating the continuous drift ZFS for bare particles in PBS. Bare particles did not equilibrate in PBS. (d-e)~Representative ZFS tracking curves during $sim$1~hr of laser excitation of core-shell particles in PBS. $N=12$ data sets were averaged to produce the right panel of Fig. \ref{fig:2}(d). (f)~Change in total PL during ZFS tracking data from (a) and (d) demonstrating the gradual drop in PL in bare particles during excitation, likely due to electron transfer from the diamond to the environment. (g)~Distribution of change in key fitting parameters from $N=9$ bare (upper panel) and $N=4$ equilibrated core-shell (lower panel) ODMR spectra before and after laser excitation, showing the large variation caused due to divergence from a perfect double Lorentzian spectra (SI note \ref{SI theoretical model} and \ref{SI ODMR fitting}). While the ODMR spectra in Fig. \ref{fig:2}(b) clearly show a change before and after laser excitation, the only consistent parameter was the change in ZFS.}
    \label{figS: PBS}
\end{figure}

\clearpage
\begin{figure} [h!]
    \includegraphics[width=1\linewidth]{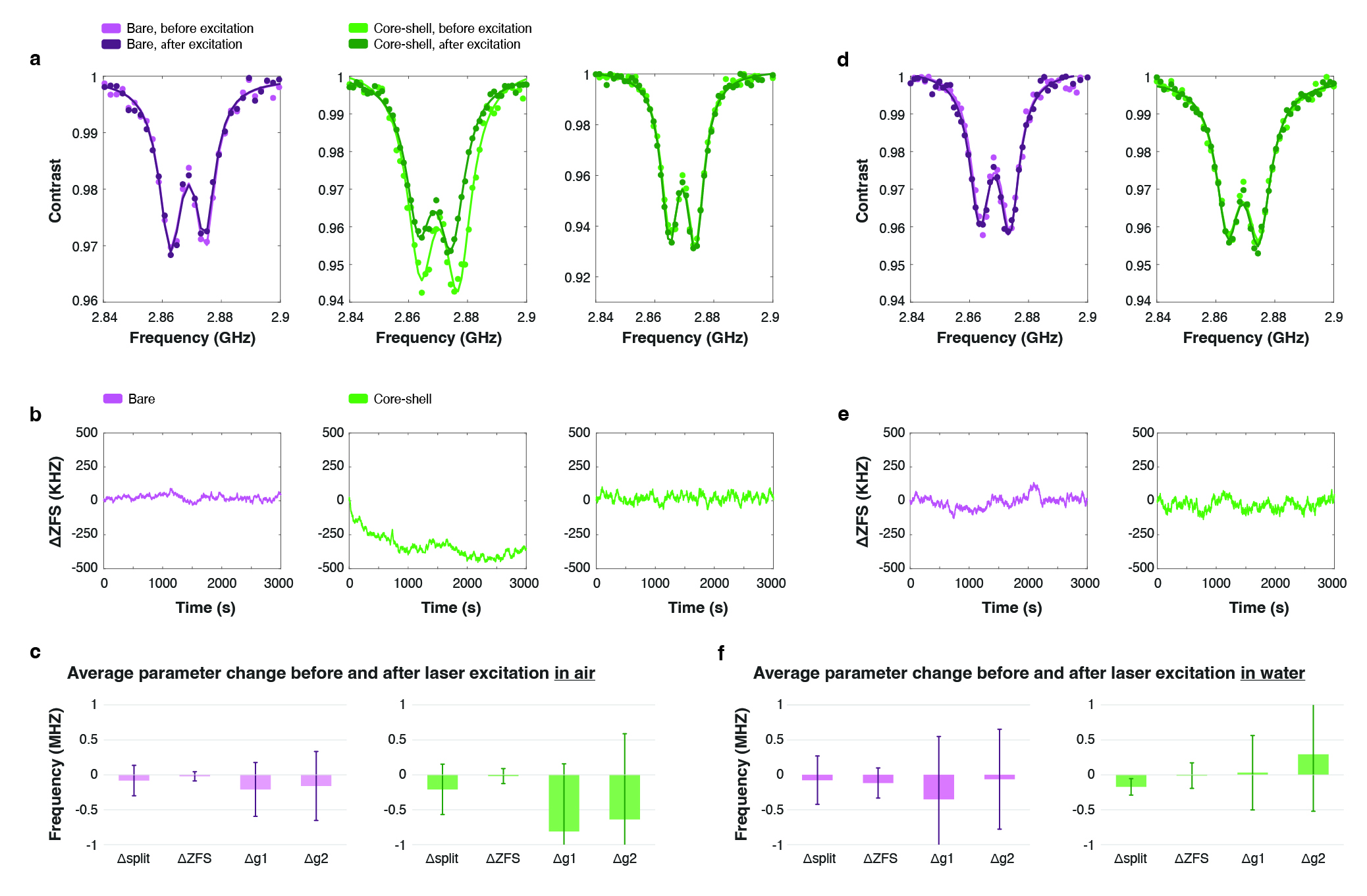}
    \caption{ZFS behavior in air and water. (a) Representative ODMR spectra before (lighter color) and after (darker color) $sim$1~hr of laser excitation in air. Bare particles showed no qualitative difference (left panel), while core-shell particles experienced a left shift towards lower frequencies (middle panel). Following a laser-power-dependent (Fig. \ref{figS: photoionization}) equilibration, core-shell particles maintained stable spectra (right panel) and no ZFS shifts were detected for several weeks (Fig. \ref{figS: photoionization}). For all spectra, points represent experimental data, while lines are double Lorentzian fits. (b) Representative ZFS tracking curves during $sim$1~hr of laser excitation in air. The time-series data for bare (left panel) pre-equilibrated core-shell (middle panel) and equilibrated core-shell (right panel) particles confirmed the trends seen in (a). (c) Distribution of change in key fitting parameters from $N=6$ bare (left panel) and $N=5$ equilibrated core-shell (right panel) ODMR spectra before and after laser excitation, showing the large variation caused due to divergence from a perfect double Lorentzian spectra (SI note \ref{SI theoretical model} and \ref{SI ODMR fitting}). Note the relative smaller variations in $\delta$ZFS. (d)~Representative ODMR spectra before (lighter color) and after (darker color) $sim$1~hr of laser excitation of bare (left panel) and core-shell (right panel) particles in MQ-water. (e)~Representative ZFS tracking curves during $sim$1~hr of laser excitation of bare (left panel) and core-shell (right panel) particles in MQ-water. (f)~Distribution of change in key fitting parameters from $N=5$ bare (left panel) and $N=4$ equilibrated core-shell (right panel) ODMR spectra before and after laser excitation.}   
    \label{figS: air and water}
\end{figure}

\clearpage
\begin{figure}[h!]
    \includegraphics[width=1\linewidth]{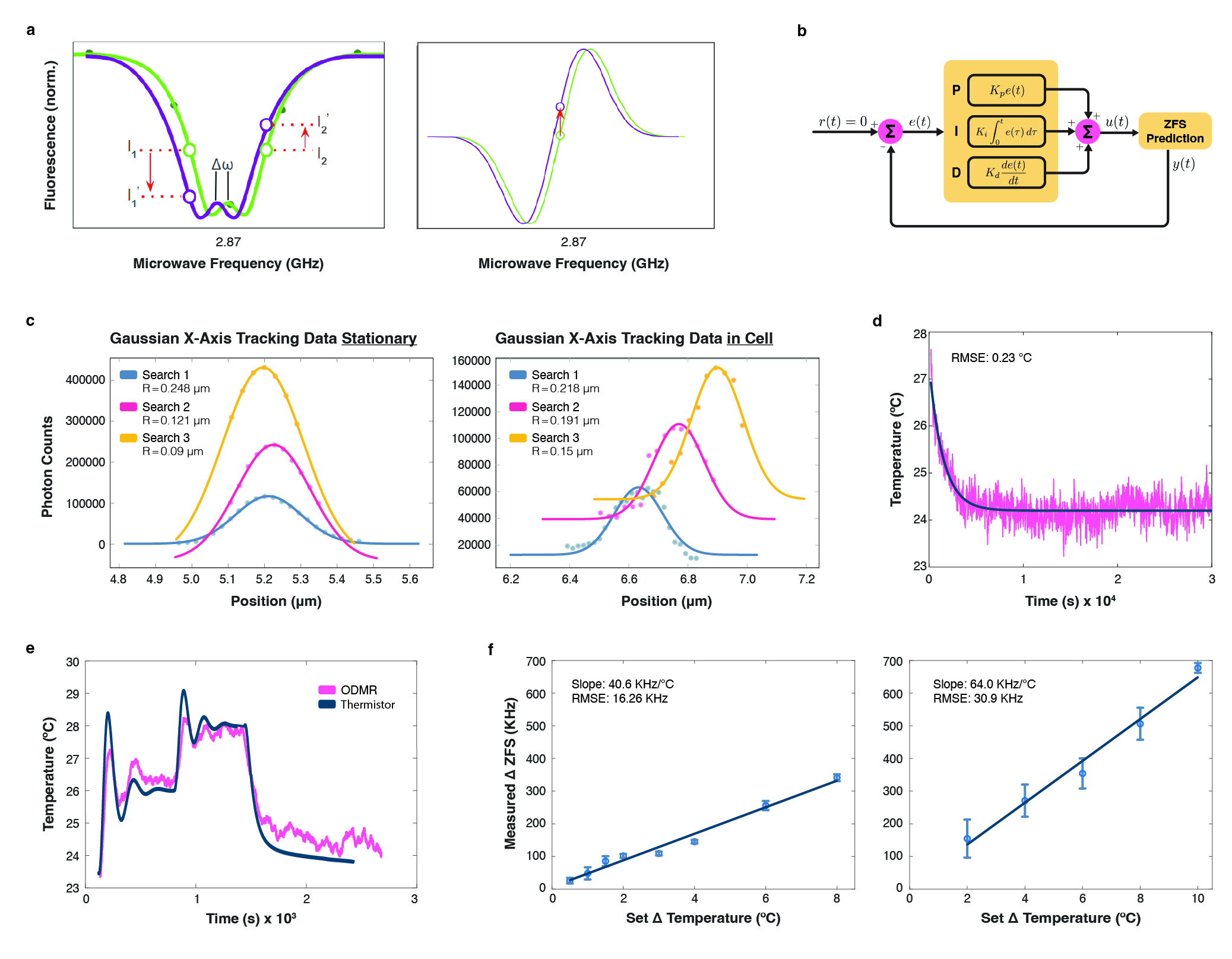}
    \caption{Methodology and performance of 2-point rapid ZFS tracking. (a)~Illustration of the changes in the two probed frequencies, $I_1$ and $I_2$, upon a left shift in the ODMR spectrum. Left panel illustrates an initial (green curve) and a left-shifted (purple curve) spectra. Subtraction of the initial spectrum from the shifted one (right panel) results in a quasi-linear regime in which the extracted quantity $I_2 - I_1$ increases and decreases upon a left or right shift, respectively. (b)~A flow diagram demonstrating the PID logic in refining our ZFS prediction. (c)~An example single-axis data from implementation of our tracking algorithm measuring a stationary (left panel) and an intracellular (right panel) single diamond nanocrystal. The search radius changes to maximize signal, decreasing for slow moving particles and increasing for faster ones (see SI note \ref{SI Rapid ZFS tracking}). (d)~ZFS tracking for a single diamond nanocrystal in air during a temperature drop. The blue line is the exponential fit \( T(x) = c \cdot e^{-\frac{x}{t}} + T_a \), where $T_a$ is the ambient temperature, $c$ is the initial amplitude of the temperature deviation from $T_a$, and $t$ is the characteristic time constant of the decay. A root-mean-square (RMS) error of \SI{0.23}{\degreeCelsius} demonstrate the high precision of this approach at a high temporal resolution of \SI{400}{\milli\second}. (e)~Simultaneous ZFS and thermistor data tracking during temperature modulation of a TEC controller demonstrating the robustness of our PID rapid ZFS tracking. The differences in temperature between the ZFS and the thermistor data are mainly a result of a distance of a few millimeter between the thermistor and the nanocrystal. (f)~Temperature dependence of a bare nanocrystal in air utilizing \SI{50}{\second} integration of our rapid ZFS tracking measurement using an oil (left panel) and an air (right panel) objectives. Oil objective measurements were limited by a larger distance between the thermistor and the nanocrystal, leading to a lower set temperature dependence, but featured lower variation and higher precision ($RMS error = 16.3$~KHz) compared with the air objective measurements ($RMS error = 30.90$~KHz). Each point is comprised of 3 different measurements with error bars representing one standard deviation.}
    \label{figS: rapid ZFS tracking}
\end{figure}

\clearpage

\begin{figure}[h!]
    \includegraphics[width=1\linewidth]{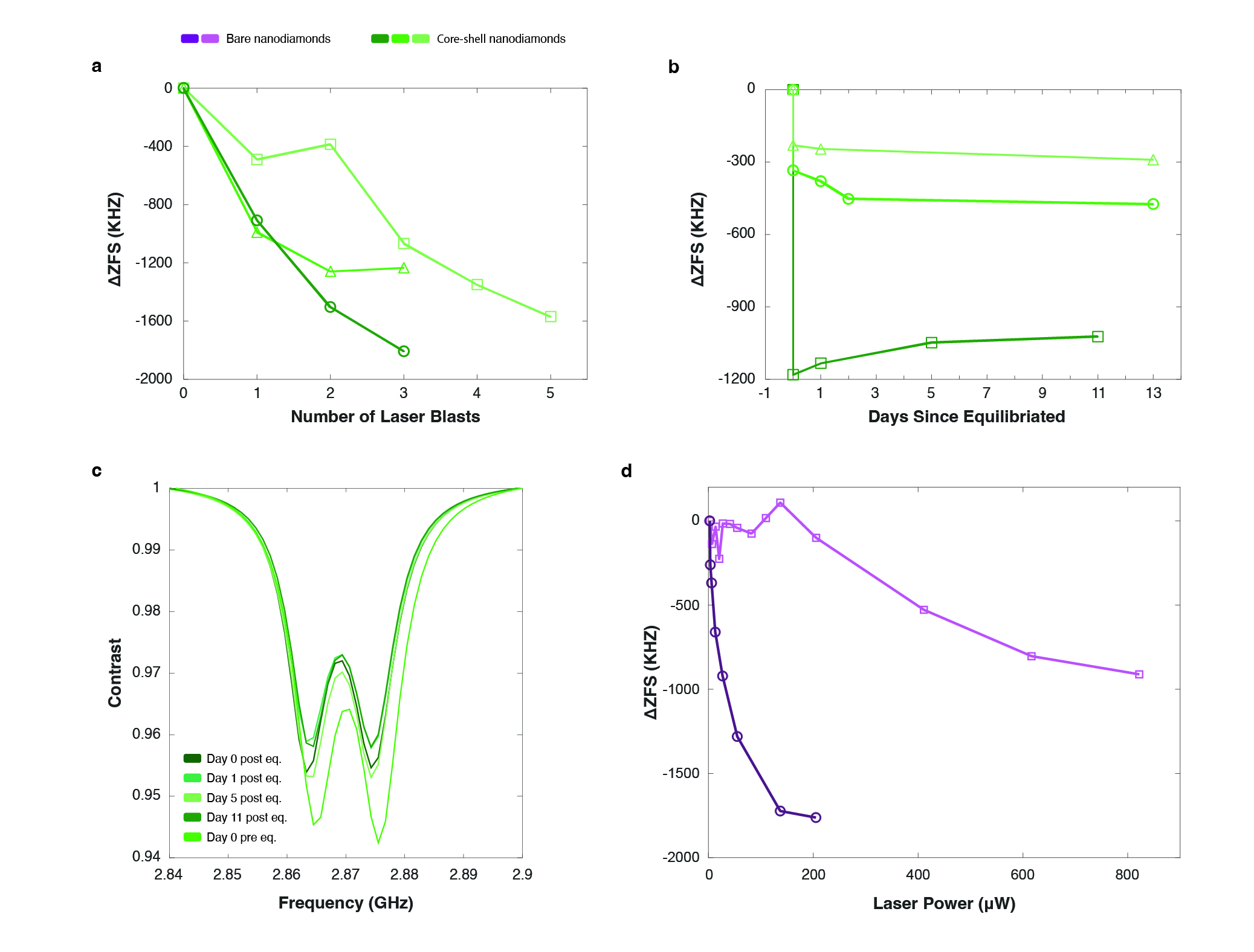}
    \caption{Further evidence for photo-induced charge transfer. (a)~ZFS shifts in three different core-shell particles following consecutive \SI{30}{\second} laser illumination pulses at higher power (\SI{55}{\micro\watt} compared with 7-\SI{10}{\micro\watt} during measurements). Shifts on the order of 1~MHz are obtained within just a few pulses, compared with the thousands of seconds time scale shown in Fig. \ref{figS: air and water}(b). Blasts at higher power achieved equilibration after a few seconds. (b)~ZFS stability of three different core-shell particles up to 13 days following equilibration. Initial point at 0~KHz represent the initial ZFS before equilibration. The spectra in (c) show a representative core=shell nanoparticle before (Day 0 pre eq.) and days after (post eq.) equilibration. The spectra clearly demonstrate the shift towards lower frequencies and its stability over time, ruling out any temperature related ZFS shifts. (d)~Increasing ZFS shifts for two bare particles in PBS obtained from ODMR spectra taken at various laser powers. The end points represent the highest power in which we were able to obtain a reasonable contrast in order to fit the spectra. We were unable to achieve equilibration of ZFS for bare particles in PBS, implying that laser illumination over time, and not just the laser power, play a role in the observed ZFS shifts.}
    \label{figS: photoionization}
\end{figure}

\clearpage

\begin{table}[ht]
    \centering
    \caption{Summary of p-values from t-tests assessing significance between groups in toxicity and inflammation experiments and in the drift term analysis for ZFS time traces in PBS and in RAW cells.}
    \begin{tabular}{|c|c|c|c|c|c|}
        \hline
        \textbf{LDH assay} && \SI{10}{\micro\gram\per{\milli\liter}} & \SI{50}{\micro\gram\per{\milli\liter}} &\SI{200}{\micro\gram\per{\milli\liter}} & \\ \hline
        24~hrs & B40 vs C40 & 0.116 & 0.292 & 0.022 & \\ \hline
        & B70 vs C70 & 0.186 & 0.050 & 0.062 & \\ \hline
        6~hrs & B40 vs C40 & 0.619 & 0.373 &  0.241 & \\ \hline
        & B70 vs C70 & 0.337 & 0.169 & 0.151 & \\ \hline
        48~hrs & B40 vs C40 & 0.076 & 0.889 & 0.066 & \\ \hline
        & B70 vs C70 & 0.030 & 0.047 & 0.062 & \\ \hline
        & & & & &\\ \hline
        \textbf{NF-$\kappa$B assay} && \SI{10}{\micro\gram\per{\milli\liter}} & \SI{50}{\micro\gram\per{\milli\liter}} & \SI{100}{\micro\gram\per{\milli\liter}} & \SI{200}{\micro\gram\per{\milli\liter}} \\ \hline
        Overnight & B40 vs C40 & 0.839 & \num{1.83e-3} & 0.010 & \\ \hline
        & B70 vs C70 & 0.205 & 0.305 & 0.099 & 0.085\\ \hline
        & B40 vs Cells & \num{3.59e-8} & \num{3.13e-13} & \num{7.49e-20} & \\ \hline
        & C40 vs Cells & \num{7.49e-9} & 0.074 & \num{1.17e-14} & \\ \hline
        & B70 vs Cells & 0.027 & 0.027 & \num{5.15e-5} & 0.021 \\ \hline
        & C70 vs Cells & 0.047 & 0.029 & 0.017 & \num{5.10e-4} \\ \hline
        & & & & &\\ \hline
        \textbf{TNF-$\alpha$ assay} && \SI{10}{\micro\gram\per{\milli\liter}} & \SI{50}{\micro\gram\per{\milli\liter}} & \SI{100}{\micro\gram\per{\milli\liter}} & \\ \hline
        Overnight & B40 vs C40 & 0.708 & 0.013 & 0.022 & \\ \hline
        & B70 vs C70 & 0.033 & \num{7.55e-3} & \num{1.56e-4} & \\ \hline
        & B40 vs Cells & 0.019 & \num{6.23e-4} & 0.013 & \\ \hline
        & C40 vs Cells & 0.139 & 0.013 & \num{9.13e-5} & \\ \hline
        & B70 vs Cells & 0.068 & 0.003 & \num{1.81e-6} & \\ \hline
        & C70 vs Cells & 0.074 & 0.021 & 0.023 & \\ \hline
        & & & & &\\ \hline
        \textbf{Drift analysis} & & & & & \\ \hline
        In PBS & Bare vs core-shell & 0.018 & & & \\ \hline
        & & & & &\\ \hline
        In RAW cells & & Core-shell nascent & Core-shell LPS & Bare nascent & Bare LPS \\ \hline
                              & Core-shell nascent & X & 0.15 & 0.84 & \num{5.74e-6} \\ \hline     
                              & Core-shell LPS & 0.15 & X & 0.58 & \num{3.85e-6} \\ \hline
                              & Bare nascent & 0.84 & 0.58 & X & \num{6.38e-3} \\ \hline           
                              & Bare LPS & \num{5.74e-6} & X & \num{6.38e-3} & \num{3.85e-6} \\ \hline     
                              
    \end{tabular}
    \label{extended_data_table1}
\end{table}	

\clearpage

\begin{figure}[h!]
    \includegraphics[width=1\linewidth]{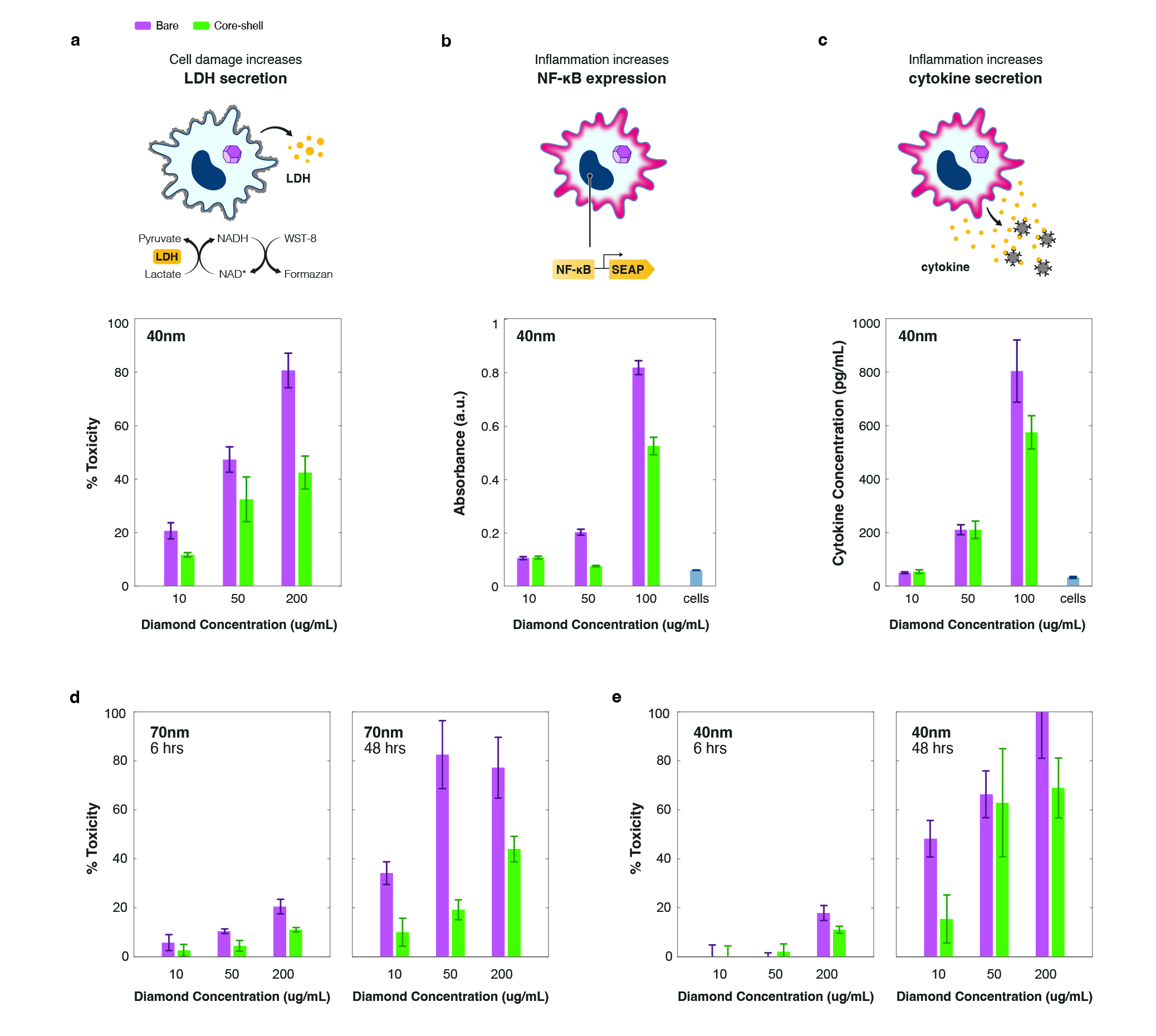}
    \caption{Additional toxicity and Inflammation data. (a-c)~LDH (left panel), NF-$\kappa$B (middle), and TNF-$\alpha$ (right panel) measurements from supernatant of RAW cells incubated with varying concentrations of 40~nm bare (purple) and core-shell (green) particles. Similar to our results with 70~nm particles, cells incubated with $\geq$\SI{50}{\micro\gram\per{\milli\liter}} core-shell nanocrystals exhibited significantly lower (although not negligible compared to untreated cells) toxicity and inflammation compared to those incubated with similar bare nanocrystals. Top illustrations depict the key measured quantities in each assay. (d)~6 (left panel) and 48 (right panel) hours time points for the LDH toxicity assay for 70~nm particles. (e)~6 (left panel) and 48 (right panel) hours time points for the LDH toxicity assay for 40~nm particles. For a list of p-values, see Extended Data Table \ref{extended_data_table1}. We note that at the 6 hours time point, 70~nm particles exhibited higher toxicity compared with 40~nm for both bare and core-shell particles. Interestingly, after overnight incubation and longer time points, the toxicity was more dominant at the 40~nm particles. This might be due to the higher number of particles per unit mass, as well as the higher surface to volume ratio for 40~nm particles that might lead to aggregations.}
    \label{figS: Toxicity and Inflammation}
\end{figure}				
								
\clearpage

\begin{figure}[h!]
    \includegraphics[width=1\linewidth]{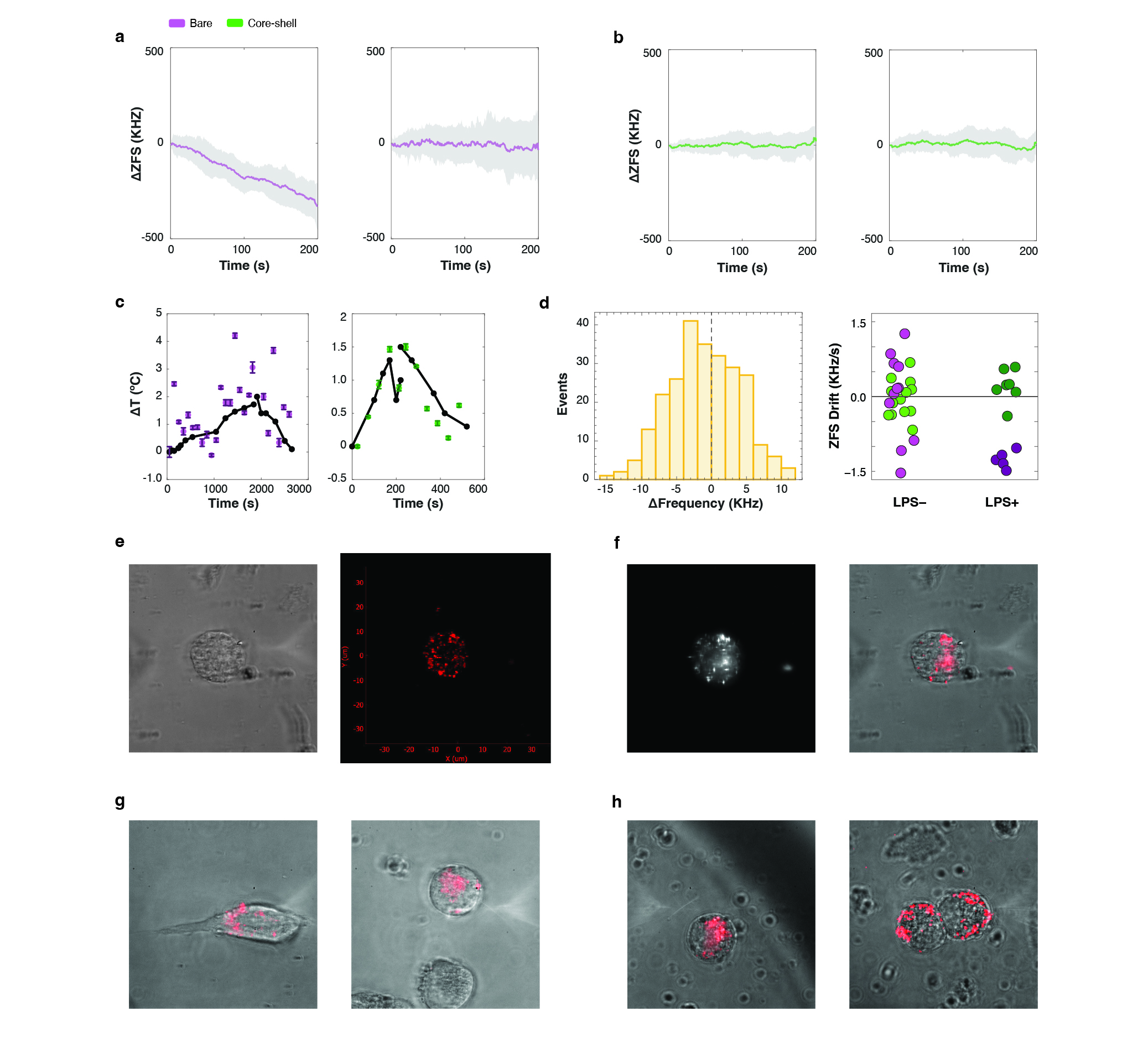}
    \caption{RAW cells ZFS measurements with bare and core-shell nanocrystals. (a)~An average of all bare particles' ZFS time traces in inflamed (left panel) and nascent (right panel) RAW cells showing a clear distinction between the groups. (b)~An average of all core-shell particles' ZFS time traces in inflamed (left panel) and nascent (right panel) RAW cells showing no qualitative change between the groups. In both (a) and (b), the gray shaded area represent one standard deviation. (c)~ZFS tracking from bare (purple, left panel) and core-shell (green, right panel) particles inside of a RAW cell. the points represent a ~\SI{100}{\second} integration during active modulation and monitoring of the temperature of the solution (see methods). As expected, the core-shell particle exhibited improved stability of ZFS and followed the modulated temperature well (RMS error of \SI{0.22}{\degreeCelsius}) compared to the bare particle (RMS error of \SI{1.25}{\degreeCelsius} from the thermistor's measured temperature. See SI note \ref{SI ZFS tracking during temperature modulation} for details about this measurement. (d)~Drift term analysis showing a representative error distribution histogram (left panel) for a bare particle ZFS time series taken in LPS stimulated RAW cells. The deviation of the mean from zero represents a drift term. The right panel shows all the drift terms for intracellular bare (purple) and core-shell (green) particles. See Extended Data Table \ref{extended_data_table1} for t-tests results. (e)~Bright field (left) and confocal (right) imaging of bare particles in a non-stimulated RAW cell. After background removal and applying thresholds, the confocal image is overlaid on the bright field image to produce Fig. \ref{fig:4}(a). An overlap (right panel) of the bright and fluorescence (left) wide-field images is shown in (f) for comparison. (g) and (h)~show two representative overlap images for intracellular bare and core-shell particles, respectively.}
    \label{figS: ZFS in RAW cells}
\end{figure}

\clearpage

\appendix
\renewcommand{\thefigure}{S\arabic{figure}}
\renewcommand{\thesection}{S\arabic{section}}
\setcounter{figure}{0} 
\setcounter{section}{0}
\setcounter{table}{0}

\section*{\huge Supplementary}

\section{Theoretical model}
\label{SI theoretical model}
We aimed to develop a comprehensive model describing the interaction of NDs with their environment under conditions relevant to NV biosensing experiments. As such, our system is comprised of type 1b, oxygen-terminated NDs hosting an ensemble of NVs that are placed under laser excitation (Fig. \ref{fig:1}(a)). Conventional NV-based electric field sensing relies on the dipole moments $d_\parallel$ and $d_\perp$, which couple to the axial and transverse electric field components, respectively\cite{Dolde2011-bu} (Fig. \ref{fig:1}(b)). In randomly oriented NV ensembles, these interactions are dependent on the relative orientation of each NV center. This orientation dependence leads to an averaging effect, resulting in ODMR line broadening\cite{Michl2019-rt} that varies randomly among different diamond nanoparticles, thus limiting their utility in sensing applications. 

To address these limitations, we consider the full ground-state Hamiltonian in the $\{\left|+1\right\rangle,\left|0\right\rangle,\left|-1\right\rangle \}$ triplet basis\cite{PhysRevB.110.024419,PhysRevB.85.205203}:
\begin{equation}
\frac{{\cal H}}{h} = \left(\begin{array}{ccc}
\frac{D}{3}+\frac{d_{\parallel}}{3}E_{z}+\gamma B_z & \frac{d'}{\sqrt{2}}E_{-}+\frac{\gamma}{\sqrt{2}}{B_{-}} & -d_{\perp}E_{+}\\
\frac{d'}{\sqrt{2}}E_{+}+\frac{\gamma}{\sqrt{2}}B_{+} & -\frac{2D}{3}-\frac{2d_{\parallel}}{3}E_{z} & -\frac{d'}{\sqrt{2}}E_{-} +\frac{\gamma}{\sqrt{2}}B_{-}\\
-d_{\perp}E_{-}  & -\frac{d'}{\sqrt{2}}E_{+} +\frac{\gamma}{\sqrt{2}}B_{+} & \frac{D}{3}+\frac{d_{\parallel}}{3}E_{z} -\gamma B_z
\end{array}\right),   
\label{hamiltonian}
\end{equation}
where $h$ is Planck's constant, $\gamma$ the gyromagnetic ratio, $\mathbf{S}=(S_x,S_y,S_z)$ the spin-1 matrices, $\mathbf{B}=(B_x,B_y,B_z)$ the magnetic field, $\mathbf{E}=(E_x,E_y,E_z)$ the electric field, $E_\pm=E_x\pm i E_y$, $B_\pm =B_x\pm i B_y$, and $D$ the temperature-sensitive zero field splitting (ZFS). Importantly, we include the commonly neglected $d'$ dipole term, which couples the $\ket{0}$ state to $\ket{\pm1}$ states (Fig. \ref{fig:1}(b)) and is estimated to be of similar magnitude as $d_{\perp}$\cite{Michl2019-rt,Policies-Practices2018-sa,Chen2020-we}. Using perturbation theory, we calculate the leading order (second-order) influence of $d'$ on the transition frequencies of our ODMR spectra
\begin{equation}
    f_{\ket{0}\rightarrow\ket{\pm}} = D + d_{\parallel}E_{z} \pm |d_{\perp}E_{\perp}| + \frac{|d'E_{\perp}|^{2}}{2D}[2\pm\cos(3\phi)],
    \label{freqODMR}
\end{equation}
with $E_\perp=\sqrt{E_{x}^2+E_{y}^2}$ and $\phi=\arctan{(E_y/E_x)}$. Considering $\textrm{ZFS}=(f_{\ket{0}\rightarrow\ket{+}}+f_{\ket{0}\rightarrow\ket{-}})/2-D$, and the ensemble-average ($\left\langle... \right\rangle$) yielding $\left\langle E_z\right\rangle\rightarrow0$, the effects of the first-order $E$-field terms cancel, limiting their implementations in electric field sensing. In contrast, the dependence of the ZFS on $d'$ yields $\left\langle \textrm{ZFS} \right\rangle= \left\langle |E_{\perp}|^{2}\right\rangle |d'|^2/D$. It is clear, therefore, that $d'$ unlocks a new possibility of sensing electric fields with an ensemble of NVs via shifts of the ZFS.    

To account for the electric field experienced by each NV, we follow the procedure described in ref.\cite{Zvi2023-zd} and solve the Poisson's equation, accounting for the implanted nitrogen/P1, vacancies, and NVs~\cite{Deak2014-hw,Shenderova2019-lm} (Fig. \ref{figS: model}). Figure \ref{fig:1}(c) describes the P1 densities (upper panel) and the resulting electric field profiles (lower panel) for a spherical particle with surface potentials $\phi_S=0.5$~V, corresponding to the expected potential on a bare, oxygen-terminated nanocrystals\cite{Zvi2023-zd}, and $\phi_S=-0.5$~V to model an environment-induced shift. 

Finally, we apply a Lindblad formalism to obtain an ODMR simulation for 100, randomly oriented NVs under laser excitation and Rabi driving. Figure \ref{fig:1}(d) illustrates the corresponding spectra for $\phi_S=0$~V and $\phi_S=0.5$~V with and without accounting for the influence of $d'$. The clear shift in the spectra confirms our analytical results in Eq. \ref{freqODMR}. To demonstrate the effect on ZFS measurements, we use our Lindblad formalism to plot the ZFS dependence on the electric field magnitude at the surface for various possible values of $d'$ (Fig. \ref{fig:1}(e)).     

\section{ODMR fitting and extraction of dipole terms}
\label{SI ODMR fitting}
ODMR fitting is done using a double Lorentzian of the form:
\begin{equation}
   f(x) = \frac{A_1}{1 + \left(\frac{2(x - x_1)}{\gamma_1}\right)^2} + \frac{A_2}{1 + \left(\frac{2(x - x_2)}{\gamma_2}\right)^2} + y_0,
    \label{doubleLorFit}
\end{equation}
where the fitting parameters $x_1$ and $x_2$ represent the $f^-$ and $f^+$ transitions, $A_1$ and $A_2$ represent their respective contrasts, $\gamma_1$ and $\gamma_2$ represent their respective FWHM, and $y_0$ represents the total PL baseline without MW modulation. This form fits well for a single NV, yet our system is comprised of a 100 NVs, each has one of eight orientations and a random position within the nanocrystal. As such, each experiences a different electric field vector and corresponding influences of the dipole moments. As captured well in our model (Fig. \ref{fig:1} and SI note \ref{SI theoretical model}), the orientation-dependent interactions with the MW AC magnetic field and band bending induced electric field give rise to different broadening, contrast, and eigenstates for each of the NVs. This results in an asymmetric PL spectrum that diverges from the double Lorentzian fit in equation \ref{doubleLorFit}. While interactions that are independent of the field's sign (i.e., the $d'$ and $d_{\perp}$ dipole moments) should not average out, the complex PL spectrum limits our ability to extract the compounded $f^-$ and $f^+$ transitions and determine the appropriate dipole terms.

Interestingly, when fitting experimental ODMR data, the obtained ZFS left shifts were consistent, while the $f^-$ to $f^+$ splits determined by $d_{\perp}$ presented high variability and no consistent trend (Fig. \ref{figS: PBS} and \ref{figS: air and water}). This is possibly due to the overall broadening caused by $d_{\perp}$\cite{Michl2019-rt}, which isn't easily distinguished from other noise sources like AC magnetic fields. We emphasize here that while our experiments where inspired by the model, the limitations in the extraction of the Lorentzian features prevents us from making any quantitative determinations of the electric field changes during our experiments. We hope that collecting additional data would allow us to refine our model (for example: including the probability to lose the NV charge state during laser excitation, and considering changes in spin $T_1$ and optical lifetime in various NVs during our measurement).  

\section{Rapid ZFS tracking}
\label{SI Rapid ZFS tracking}

\subsection*{2-points measurement}
We employ a 2-point measurement scheme \cite{Singam_2020} to estimate the zero-field splitting (ZFS) frequency, $\omega_{\mathrm{ZFS}}$ of our NV center over time. In this scheme, given a center frequency $\omega_c$, we measure the fluorescence at two microwave frequencies, $\omega_1 = \omega_c - \frac{\gamma}{2}$ and $\omega_2 = \omega_c + \frac{\gamma}{2}$, where $\gamma$ is the full-width half maximum of the optically detected magnetic resonance (ODMR) spectrum of the NV. The fluorescence values at these frequencies are labeled $I_1$ and $I_2$. The choices of $\omega_1$ and $\omega_2$ are in the quasi-linear regime of the ODMR Lorentzian spectrum so that changes in the values of $I_1$ and $I_2$ correspond linearly with changes in $\omega_{\mathrm{ZFS}}$. To utilize this property, we plot the "frequency jump" function, $f(\omega_c) = I_2(\omega_c)-I_1(\omega_c)$ (this function evaluates to $0$ at $\omega_c=\omega_{\mathrm{ZFS}}$ due to symmetry to the Lorentzian). The slope of this function near its zero is quasi-linear due to the aforementioned properties of the Lorentzian, and so for any given deviation in $I_2-I_1$ reading from $0$, we can compute the change in ZFS, $\omega_{\mathrm{ZFS}}$, as 
\begin{equation}
    \Delta\omega_{\mathrm{ZFS}} = -\frac{I_2-I_1}{\alpha},
    \label{quasilinear}
\end{equation}
where $\alpha$ is the quasilinear slope at the zero point. 

Once the quasilinear slope has been characterized, we implement sideband modulation to enable rapid switching between $\omega_1$ and $\omega_2$, centered around the initial $\omega_{\mathrm{ZFS}}$. Rather than directly measuring \(\Delta\omega_{\mathrm{ZFS}}\) as outlined in Eq.~\ref{quasilinear}, we input this value as a "process variable" into a PID control loop to mitigate noise and fluctuations in the readings. In this setup, the set point error for the ZFS reading is defined as \(u(t) = 0\), the predicted \(\Delta\omega_{\mathrm{ZFS}}\) value (process variable) is represented by \(y(t)\), and the error fed into the PID computation is \(e(t) = -y(t)\). This feedback loop effectively stabilizes the predicted \(\Delta\omega_{\mathrm{ZFS}}\) to have a $0$ error, reducing jitter and suppressing the effects of noisy readouts. Running this process continuously over our measurement, we generate estimated $\omega_{\mathrm{ZFS}}$ values over time.

\subsection*{Spatial Tracking of the NV}
Fluorescence measurements of the NV center are highly sensitive to the focus of the confocal microscope. In dynamic environments, such as those involving fluctuating temperatures and moving fluids (e.g., within cells), it is necessary to implement a tracking system capable of concurrently acquiring fluorescence data for ZFS prediction while monitoring the NV center's position over time to refocus the measurement stage. To do this, we perform a sweep along the three spatial axes, centered at the previous focus position, with a sweep radius $R$. For example, along the $x$-axis, the sweep spans the interval $[x_0-R, x_0+R]$, where $x_0$ is the previous $x$-axis focus position. At each point along the sweep, we measure the fluorescence counts $I_{1,2}$. The fluorescence profile of the NV center, modeled as a point light source, exhibits a Gaussian distribution. Therefore, the total fluorescence collected at each point is used to estimate the new center position of the NV along the corresponding axis by fitting the measured fluorescence profile to a Gaussian function. Meanwhile, since the $I_{1,2}$ measurements are taken over equal durations at each position, we can sum them up individually and use them to compute \(\omega_{\mathrm{ZFS}}\).  

Furthermore, we implemented a PID control loop to lock the sweep radius $R$, maintaining it at twice the drift length of the NV center measured in the previous observation along each spatial axis. Fig. \ref{figS: rapid ZFS tracking} features examples of $x$-axis sweeps showing an increase in PL as the sweep-radius decreases showing convergence on the position of a stationary NV. In contrast, when the NV center is moving within a cell, the fluorescence data reveals a shift in the $x$-coordinate of the NV across multiple sweeps. Although $R$ decreases in this case as well, it does not reach as small a value as in the stationary scenario due to the motion of the NV. Sweeps are performed sequentially along all three axes, with the coordinates of the other two axes held constant while sweeping along one axis, ensuring precise localization along each dimension.

\section{Evidence for photo-assisted charge transfer}
\label{SI Evidence for photo-assisted charge transfer}
While the charge transfer between the diamond and the environment is linked to the band bending profile, the existence of non resonant laser excitation may promote this affect, and even lead to an ionization of defect states that otherwise would remained occupied. The suppression of the electron transfer process in core-shell particles is likely a combination of the protection of the surface from the environment and the energy barrier that reduces the probability of electron tunneling. Such an effect should be scale exponentially with the thickness of the shell and should be further explored. While we cannot conclusively determine that laser excitation is necessary for the charge transfer to happen in the conditions measured here, our findings suggest the the reported ZFS shifts are promoted by photo-excitation. Fig. \ref{figS: photoionization} shows several supporting evidence, including the laser-power dependence of the ZFS shifts of bare particles in PBS (Fig. \ref{figS: photoionization}(d)), as well as the effects of high power laser exposure on the initial equilibration of core-shell particles (Fig. \ref{figS: photoionization}(a-b)). Particularly interesting is the lower frequency shift experienced by core-shell particles during equilibration, which suggests that electrons might initially flow into the diamond during this process, as the positive electric field is decreasing. This is in contrast to bare particles, which likely experience an electron flow out of the diamond as their electric field turns less negative. This is further supported by the PL profiles in Fig. \ref{figS: PBS}, showing the stability in PL in core-shell particles vs the decrease in bare particles with increased duration of illumination. 

\section{Toxicity and Inflammation}
\label{SI Toxicity and Inflammation}

\subsection*{LDH assay}
Following incubation with bare or core-shell particles and transfer of \SI{50}{\micro\liter} of supernatant to a new flat-bottomed 96 well plate, \SI{50}{\micro\liter} of Reaction Mixture (kit Cat. No. C20300) was added to each sample well and mixed well. The mixture was shaken in the dark at room temperature for 30 minutes. Subsequently, \SI{50}{\micro\liter} of Stop Solution were added to each well followed by mixing. Absorbance was measured at \SI{490}{\nano\meter} from which the absorbance at \SI{680}{\nano\meter} was subtracted. For maximum and minimum LDH activity, we incubated cells with \SI{10}{\micro\liter} 10X Lysis buffer or \SI{10}{\micro\liter} sterile water, respectively, instead of incubation with NDs. We calculated $\%\mathrm{Toxicity} = \frac{\mathrm{Sample \ activity - spontaneous\ activity}}{\mathrm{Max\ activity - spontaneous\ activity}}$.

\subsection*{NF-$\kappa$B assay}
Following incubation with bare or core-shell particles and transfer of \SI{20}{\micro\liter} of supernatant to a new flat-bottomed 96 well plate, \SI{180}{\micro\liter} of QUANTI-Blue™ solution was added to each sample well and mixed well. The mixture was incubated at 37°C for 3 hours, and absorbance was measured at \SI{620}{\nano\meter} from which the absorbance at \SI{680}{\nano\meter} was subtracted. As control, absorbance was measured from cells with no diamond nanocrystals.

\subsection*{Inflammatory cytokine secretion}
Murine RAW Blue™ macrophages were incubated with bare or core-shell particles for 12 hours (overnight). The resulting supernatant was analyzed for cytokine secretion using a LEGENDplex™ Mouse Inflammation Panel (13-plex) kit (BioLegend, Cat. No. 740150) according to the manufacturer’s instructions. In brief, standard curves for cytokine concentrations were created using the manufacturer’s Standard Cocktail, which was diluted as instructed. For each experimental well, \SI{15}{\micro\liter} of supernatant were added to the provided 96-well V-bottom plate and mixed with \SI{10}{\micro\liter} of Assay Buffer. Subsequently, \SI{25}{\micro\liter} of Capture Beads and additional Assay Buffer were added to both standard and experimental wells. The plate was sealed, protected from light, and incubated with shaking at 700 rpm for two hours. Following the initial incubation period, the plate was centrifuged at 1050 rpm for five minutes. The resulting supernatant was removed, and the remaining beads were washed with \SI{200}{\micro\liter} of 1X Wash Buffer. Centrifugation was repeated at 1050 rpm, and the Wash Buffer was removed. Next, \SI{25}{\micro\liter} of Detection Antibody solution were added to each well. The plate was sealed, protected from light, and incubated with shaking at 700 rpm for one hour, after which \SI{25}{\micro\liter} of SA-PE were added directly to each well without washing. The plate was sealed, covered, and incubated for an additional 30 minutes with shaking at 700 rpm. The plate was then centrifuged at 1050 rpm, and the beads were washed with \SI{200}{\micro\liter} of 1X Wash Buffer. Centrifugation was repeated at 1050 rpm, and the Wash Buffer was carefully removed. Finally, the beads were reconstituted in \SI{150}{\micro\liter} of Wash Buffer, and the plate was analyzed using a flow cytometer (Agilent ACEA NovoCyte® 1000).

\section{Thermodynamic analysis of RAW cell thermogenesis}
\label{SI Thermodynamic analysis of RAW cell thermogenesis}
To verify that the observed 268$\pm$\SI{29}{\kilo\hertz} ZFS shift in bare nanodiamonds does not stem from an actual temperature increase, we employ thermodynamic considerations to demonstrate that such a variation is not physically attainable within a cellular environment. We follow the dimensional analysis approach for steady-state conditions in a dense medium, as presented in~\cite{baffou2014}, which states that to the correct order of magnitude:
\[
    \Delta T \sim \frac{P}{\kappa L}
\]
Here, \(\Delta T\) represents the temperature increase at a given site, \(P\) is the input power from the heat source, \(L\) is the characteristic scale of the heat source, and \(\kappa\) is the thermal conductivity of the surrounding medium. Assuming an aqueous environment, $\kappa \sim \SI{1}{\watt \per\meter \per\kelvin}$. The observed ZFS shift corresponds to a temperature change $\Delta T \sim \SI{4}{\kelvin}$ over $\SI{200}{\second}$. Finally, to estimate the minimum power required for inducing a temperature change, we consider the most favorable scenario by selecting the smallest cellular structure capable of generating heat—the mitochondrion with $L \sim \SI{1}{\micro\meter}$—as the heat source. This corresponds to a power of $P \sim \SI{20}{\nano \watt}$ delivered in the cell by the heat source. However, the typical thermal power generated by a cell is approximately $\sim \SI{100}{\pico\joule}$ \cite{LOESBERG}, making the value estimated here implausible, as it exceeds observed values by two orders of magnitude. As such, we exclude the possibility of the zero-field splitting (ZFS) shift arising from temperature-changing processes within the cell.

\section{ZFS tracking during temperature modulation}
\label{SI ZFS tracking during temperature modulation}
While no systematic shifts were detected in nascent cells, the substantial fluctuations present a major challenge to thermometry applications. We demonstrate the challenge of decoupling temperature measurements from the effects reported here by manipulating the temperature of the stage-top chamber using a high-precision resistive heater (Thorlabs Inc., model HT19R). We simultaneously tracked the coverslip temperature and the ZFS of a bare and a core-shell particle in nascent RAW cells (Fig.~\ref{figS: ZFS in RAW cells}(c)). We determined the measured temperature by integrating 100 seconds of ZFS tracking data. Core-shell particles followed the temperature well with average standard deviation of \SI{0.25}{\degreeCelsius}, with an RMS error of \SI{0.22}{\degreeCelsius} from the thermistor's measured temperature. In contrast, bare particles presented an average standard deviation of \SI{1.12}{\degreeCelsius} and a significantly larger RMS of \SI{1.25}{\degreeCelsius} from the thermistor's measured temperature.

\section{Coplanar waveguide}
\label{SI Coplanar waveguide}
In experiments involving cells, microwaves were delivered using a lithographically fabricated coplanar waveguide. The waveguide is fabricated by defining a waveguide design on a 200 $\mu$m thick glass microscope coverslip. Following the lithographic patterning, \SI{50}{\nano\meter} Ti, \SI{1200}{\nano\meter} Cu, and \SI{100}{\nano\meter} Au are deposited on the top before lifting off the underlying photoresist to create an $\Omega$ shaped antenna with a line width of \SI{20}{\micro\meter} and a curvature radius of \SI{200}{\micro\meter}. The waveguide is designed to be impedance matched (50 $\Omega$) and to deliver microwaves to the opposite (clean glass with no metal) side of the coverslip. This is done to prevent metal contamination of the cells and to create a large uniform area of microwave field. To perform the experiments, the coplanar waveguides are mounted on to custom-made printed circuit boards (PCB) using a low temperature indium-based solder.

\section{Core-Shell synthesize}
\label{SI Core-Shell synthesize}
The growth of Silica shells on diamond nanocrystals was performed using a previously described process utilizing a sol-gel Stöber process\cite{Stober1968-uz} with a tetraethyl orthosilicate (TEOS) precursor. A 1 mg/mL of as-purchased carboxylated 70~nm diamond nanocrystals were sonicated for 20~min. Meanwhile, \SI{8}{\milli\gram} PVP (\SI{10}{\kilo\dalton}; Sigma-Alderich) was dissolved in \SI{16.5}{\milli\liter} reversed-osmosis purified H$_2$O (MQ-H$_2$O) and sonicated for 10~min. The nanocrystals were added into the PVP solution and stirred at 600~rpm overnight. The synthesized PVP-nanocrystals were centrifuged at 20000~X~g for 30~min and the particles were redissolved by sonicated in \SI{3.75}{\milli\liter} ethanol for 20~min (at this point the solution could be stored in 4C for further use). The solution was stirred and 
\SI{8.5}{\micro\liter} of TEOS (Sigma-Alderich) was added, followed by \SI{40}{\micro\liter} of 30$\%$ ammonia. The solution was left to stir 6-10 hours after which it was purified by centrifugation for 15~min at 15000~X~g and washed twice (10~min at 15000~X~g) and finally suspended with \SI{1}{\milli\liter} ethanol. For long-term storage, the solution was placed in Acetone instead.

\section{RAW cells}
\label{SI RAW cells}
RAW-Blue™ Cells(InvivoGen), an NF-$\kappa$B-SEAP reporter Cell line derived from the murine RAW 264.7 macrophages, were cultured in DMEM (DMEM, high glucose, pyruvate Gibco 11995073) supplemented with 10\% (v/v) heat-inactivated fetal bovine serum (Gibco 16140071), and Antibiotic-Antimycotic (1X) from gibco (\SI{100}{{\units}\per{\milli\liter}} of penicillin, \SI{100}{{\micro\gram}\per{\milli\liter}} of streptomycin, and \SI{0.025}{{\micro\gram}\per{\milli\liter}} of Gibco Amphotericin B), with \SI{100}{{\micro\gram}\per{\milli\liter}} Normocin and \SI{100}{{\micro\gram}\per{\milli\liter}} selection antibiotic  Zeocin. The cells were kept in a humidified 5\% CO$_2$ atmosphere at 37°C. 

\section{Imaging system}
\label{SI Imaging system}
NV measurements were performed using a home-built confocal microscope. Optical excitation was provided by a \SI{520}{\nano\meter} pulsed laser (LABS electronics DLnsec) and focused onto the sample (confocal mode) using a X60, NA=1.49 oil-immersion objective (Olympus APON60XOTIRFl). Translation of the excitation beam was done using a fast steering mirror (Newport FSM-300). Epifluorescence emission was separated from the excitation beam using a dichroic filter (Chroma T610lpxr) and filtered (Semrock LP01-594R-25) before being focused onto a single-photon counting module (Excelitas, SPCM-ARQH-14). For cell imaging, we used white light and a \SI{488}{\nano\meter} (Cobolt 06-MLD \SI{100}{\milli\watt}) in a wide-field mode for cells and NDs, respectively. The system was equipped with a live-cell chamber (Invivo Scientific, STEV.ECU.HC5 STAGE TOP) for ZFS measurements in live cells at \SI{37}{\degreeCelsius}. 

\section{Statistics}
\label{SI statistics}
\subsection*{Allan Variance Analysis}
To compute Allan variances for individual NV ZFS time-series data, we recorded the raw fluorescence data for $I_1$ and $I_2$ and artificially generated data sets for measurements with larger averaging time by summing over subsequent fluorescence data. For example, if our initial data averaged fluorescent data measurements for $\SI{1}{second}$, then we would generate artificial data for $\SI{2}{second}$ of measurement by summing the $I_1$ and $I_2$ data from both measurements, respectively. Using the fluorescence data for each averaging time, we generated time-series data of the ZFS in the process described in Sec.~\ref{SI Rapid ZFS tracking}, and could then compute the $\tau=0$ Allan variance for each dataset. Piecing these values together, the $\tau$ axis is the averaging time and the Allan variance axis is the starting Allan variance value for each averaging time.

To compute the average Allan variance plot for the time-series ZFS data of bare (n = 8) and core-shell (n = 12) nanodiamonds in PBS, we applied a weighted average of the individual Allan variance plots with the weights being the total measurement duration. This method ensures that the aggregate plot accurately reflects the contribution of each time series based on its respective measurement period. We then obtained Allan variance values of $6.90$ for the bare nanodiamonds and $4.59$ for the core-shell nanodiamonds. The minimum Allan variance recorded was $1.70$ for the bare nanodiamonds and trended below $0.82$ for the core-shell nanodiamonds, though the average Allan variance plot for core-shell did not exhibit a clear minimum point. These findings suggest that in PBS, the core-shell diamond nanocrystals ZFS is more stable against both low and high-frequency noise sources than the bare diamond nanocrystals ZFS.

\subsection*{Augmented Dickey-Fuller Test}
The Augmented Dickey-Fuller (ADF) Test is a statistical test used to determine whether a given time series is stationary or contains a unit root, which would indicate non-stationarity. It includes lagged differences of the time series to account for autocorrelation in the data. The ADF test estimates the following regression equation:

\[
\Delta y_t = \alpha + \gamma y_{t-1} + \sum_{i=1}^{p} \delta_i \Delta y_{t-i} + \epsilon_t
\]

Where:
\begin{itemize}
    \item \(\Delta y_t = y_t - y_{t-1}\): The first difference of the time series, capturing the change in values.
    \item \(\alpha\): Constant or intercept term
    \item \(\gamma y_{t-1}\): Coefficient on the lagged level of the series, which is tested for significance.
    \item \(\sum_{i=1}^{p} \delta_i \Delta y_{t-i}\): Lagged differences of the series, included to control for autocorrelation.
    \item \(\epsilon_t\): White noise error term.
\end{itemize}

The key test statistic is associated with \(\gamma\), which indicates the presence of a unit root. The null hypothesis (\(H_0\)) is that the series has a unit root (\(\gamma = 0\)), meaning it is non-stationary. The alternative hypothesis (\(H_1\)) is that the series does not have a unit root (\(\gamma < 0\)), meaning it is stationary.

In the ADF test, the inclusion of lagged differences (\(\Delta y_{t-i}\)) is critical for addressing autocorrelation in the series. However, choosing the right number of lags (\(p\)) is non-trivial. To automate this process, the ADF test can use the Akaike Information Criterion (AIC), which evaluates models with different lag lengths and selects the one that minimizes the AIC value.

The AIC is defined as:
\[
\text{AIC} = 2k - 2\ln(L),
\]
where:
\begin{itemize}
    \item \(k\): Number of estimated parameters (including the intercept and lag terms).
    \item \(\ln(L)\): Log-likelihood of the model.
\end{itemize}

Using AIC ensures a balance between model fit and complexity, selecting the optimal lag length to avoid overfitting or underfitting.

We use the \textit{statsmodels.tsa.stattools.adfuller} Python package to conduct the ADF test on our time-series data, and implement a threshold of a p-value \(< 0.05\) to indicate the rejection of \(H_0\), suggesting the series is stationary. A p-value of 0.557 indicates that the ZFS time series for bare nanocrystals in PBS is a non-stationary process, whereas, for core-shell particles, a p-value of \num{3.06e-4} suggests a stationary process.

\clearpage

\end{document}